\documentclass[english,12pt,a4paper]{article}
\pdfoutput=1

\usepackage[T1]{fontenc}
\usepackage[utf8]{inputenc}
\usepackage[english]{babel}
\usepackage{textcomp}
\usepackage{authblk}     
\usepackage[export]{adjustbox}% authors/affiliations
\usepackage[a4paper,margin=2cm]{geometry}
\usepackage{graphicx}
\usepackage{amsmath,amssymb}
\usepackage[dvipsnames]{xcolor}
\usepackage[normalem]{ulem}
\usepackage{enumitem}
\usepackage[mathscr]{euscript}
\usepackage{csquotes}
\usepackage[colorlinks=true,linkcolor=MidnightBlue,citecolor=MidnightBlue,urlcolor=MidnightBlue]{hyperref}
\usepackage{cite}

\title{\vspace*{-2.2cm}\rightline{\normalfont\small FERMILAB-PUB-26-0822-T}\vspace{0.6cm}{ \LARGE 
\bfseries Active Galactic Nuclei as Beyond the Standard Model Laboratories: a White Paper}}

\author[1,2]{Gonzalo Herrera\thanks{\href{mailto:gonzaloh@mit.edu}{gonzaloh@mit.edu}}}
\author[3,4]{Filippo Sala\thanks{\href{mailto:f.sala@unibo.it}{f.sala@unibo.it}, FS is on leave of absence from LPTHE, CNRS and Sorbonne U., Paris}}
\author[5,6,7]{Piero Ullio \thanks{\href{mailto:ullio@sissa.it}{ullio@sissa.it}}}
\author[8]{Misty C. Bentz}
\author[9,10,11]{\\ Carlos Blanco}
\author[5,6]{Roberto Caiozzo}
\author[12,13]{Marina Cerme\~no}
\author[14]{Matteo Cerruti}
\author[15]{James M. Cline}
\author[3,4]{Andrea Giovanni De Marchi}
\author[16,17,18]{Pierfrancesco Di Cintio}
\author[19]{Alessandro Granelli}
\author[20,21]{R. Andrew Gustafson}
\author[22]{Bradley J. Kavanagh}
\author[23]{\\ Cristina Lagunas Gualda}
\author[9,24,25]{Kohta Murase}
\author[5,6,7]{Takuya Okawa}
\author[14]{\\ Xavier Rodrigues}
\author[26,7]{Lorenzo Ubaldi}
\author[27]{Gabrijela Zaharijas}

\affil[1]{Physics Department and Kavli Institute for Astrophysics and Space Research, Massachusetts Institute of Technology, Cambridge, MA 02139, USA}
\affil[2]{Harvard University, Department of Physics and Laboratory for Particle Physics and Cosmology, Cambridge, MA 02138, USA}
\affil[3]{Dipartimento di Fisica e Astronomia, Universit\`a di Bologna, via Irnerio 46, 40126 Bologna, Italy}
\affil[4]{INFN, Sezione di Bologna, viale Berti Pichat 6/2, 40127, Bologna, Italy}
\affil[5]{Scuola Internazionale Superiore di Studi Avanzati (SISSA), via Bonomea 265, 34136 Trieste, Italy}
\affil[6]{INFN, Sezione di Trieste, via Valerio 2, 34127 Trieste, Italy}
\affil[7]{Institute for Fundamental Physics of the Universe (IFPU), via Beirut 2, 34151 Trieste, Italy}
\affil[8]{Department of Physics and Astronomy, Georgia State University, Atlanta, GA 30303, USA}
\affil[9]{Department of Physics; Department of Astronomy \& Astrophysics, Institute for Gravitation and the Cosmos, The Pennsylvania State University, University Park, PA 16802, USA}
\affil[10]{Department of Physics, Princeton University, Princeton, NJ 08544, USA}
\affil[11]{Stockholm University and The Oskar Klein Centre for Cosmoparticle Physics, Alba Nova, 10691 Stockholm, Sweden}
\affil[12]{Departamento de FAIAN, ETSIN, Universidad Politécnica de Madrid, Av. de la Memoria~4, E-28040, Madrid, Spain}
\affil[13]{Instituto de F\'isica Te\'orica UAM-CSIC, C/ Nicol\'as Cabrera, 13-15, 28049 Madrid, Spain}
\affil[14]{Université Paris Cité, CNRS, Astroparticule et Cosmologie, F-75013 Paris, France}
\affil[15]{McGill University, Department of Physics \& Trottier Space Institute, 3600 Rue University, Montr\'eal, QC, H3A 2T8, Canada}
\affil[16]{Consiglio Nazionale delle Ricerche, Istituto dei Sistemi Complessi, via Madonna del Piano 17, 50022 Sesto Fiorentino (FI), Italy}
\affil[17]{INFN -- Sezione di Firenze, via G. Sansone 1, 50022 Sesto Fiorentino (FI), Italy}
\affil[18]{INAF -- Osservatorio Astrofisico di Arcetri, Largo E. Fermi 5, 50125 Firenze, Italy}
\affil[19]{Instituto de Física Corpuscular (IFIC), CSIC-Universitat de València, Parc Científic UV, c/ Catedrático José Beltrán 2, E-46980 Paterna, Spain}
\affil[20]{Center for Neutrino Physics, Department of Physics, Virginia Tech, Blacksburg, VA 24061, USA}
\affil[21]{TRIUMF, Vancouver, BC V6T 2A3, Canada}
\affil[22]{Instituto de F\'{i}sica de Cantabria (IFCA, UC-CSIC), Av. de Los Castros s/n, 39005 Santander, Spain}
\affil[23]{Friedrich-Alexander-Universität Erlangen-Nürnberg, Hugenottenpl. 6, 91054 Erlangen, Germany}
\affil[24]{Department of Physics, The University of Tokyo, Bunkyo, Tokyo 113-0033, Japan}
\affil[25]{Center for Gravitational Physics and Quantum Information, Yukawa Institute for Theoretical Physics, Kyoto, Kyoto 606-8502 Japan}
\affil[26]{Jo\v{z}ef Stefan Institute, Jamova 39, 1000 Ljubljana, Slovenia}
\affil[27]{Center for Astrophysics and Cosmology, University of Nova Gorica, Vipavska 11c, 5270, Ajdovščina, Slovenia}
\date{} % no date

\begin{document}
\maketitle

\begin{abstract}
Multi-messenger and multi-wavelength observations of Active Galactic Nuclei (AGN) can provide otherwise inaccessible knowledge about physics Beyond the Standard Model (BSM), and may even lead to its discovery.
Realizing this potential hinges on two parallel efforts: advancing our understanding of AGN and of the dark matter around them, and identifying the BSM signatures that are at once promising and robust against the gaps in that understanding.
These fields have attracted growing interest in recent years, also motivated by the IceCube association of neutrinos to individual AGN, yet we have only begun to explore their cross-fertilization. This white paper offers a new perspective on this interdisciplinary endeavor. It concisely reviews the state of the art in each field and at their interface, and identifies the key open questions that must be addressed to strengthen our prospects for learning about BSM physics with AGN.
\end{abstract}

\tableofcontents

\newpage 
\small{
\section{Introduction}

Nature keeps hiding the answers to several of the deepest open questions in fundamental physics and cosmology: foremost among them the nature and non-gravitational interactions of dark matter (DM)~\cite{Cirelli:2024ssz}, and, more broadly, what lies beyond the smallest length and time scales that we have tested. 
Astrophysical and cosmological observations have recently taken a leading role in driving progress on these questions.
The unprecedented wealth of existing and upcoming data,  ranging from electromagnetic radiation and neutrinos to gravitational waves, 
together with their combination into multi-messenger and multi-wavelength campaigns,
offers a plethora of novel opportunities to learn about physics beyond the Standard Model (BSM).
These thrilling opportunities come, however, with significant intrinsic uncertainties and unknowns: unlike the systems studied at particle colliders and in other laboratory experiments, the initial and present states of astrophysical sources are under much poorer control.

The recent past has proven that some of these uncertainties can be overcome with a combination of more data and modeling efforts, to the point where sky observations yield solid and lasting lessons about fundamental physics. Remarkable examples include microwave photons and large-scale structures, which established cosmology as a precision science; photons, cosmic rays and neutrinos, together with galactic structures, which taught us incontrovertible lessons about the microphysics of DM that are inaccessible to laboratory experiments.
If these are now established and extremely rich avenues for fundamental physics, it is because in their early days the community was not daunted by the many associated uncertainties and unknowns.

This is the vision that motivates the new perspective advocated in this white paper: we call for a community effort towards establishing Active Galactic Nuclei (AGN) as laboratories for fundamental-physics discoveries.

\medskip

AGN~\cite{Padovani:2017zpf} are extremely luminous central regions of a galaxy powered by a supermassive black hole (BH) that is actively accreting matter,
falling toward the black hole 
and forming an accretion disc~\cite{Abramowicz:2011xu}. 

In some AGN, the accretion of matter onto the BH is accompanied by the launching of a pair of powerful, relativistic jets of plasma perpendicular to the disc, releasing enormous amounts of energy across the electromagnetic spectrum, from radio waves to X-rays and gamma-rays. 
AGN with one of the jets pointing toward Earth are called blazars.
Beyond photons, AGN have also been identified as potential sources of high-energy neutrinos \cite{IceCube:2018dnn, IceCube:2022der}, making them key targets for multi-messenger astronomy. 
The discovery of high-energy cosmic neutrinos is the landmark achievement that has recently been honored by the 2026 Nobel Prize in Physics, and neutrino observations of AGN will turn multi-messenger AGN studies into an active BSM research direction.

AGN offer opportunities to test fundamental physics that are, in principle, unique. This is due both to the high particle energies and radiation intensities reached in these systems, unmatched by other astrophysical environments and laboratory experiments on Earth, and to the extreme environments that host them, where the density of Standard Model (SM) and DM particles can be huge.
AGN then allow us to explore BSM at once along complementary and synergic routes, analogous to the “high-intensity” and “high-energy” frontiers of collider physics.
The exploration carried out so far about AGN for fundamental physics confirms the very high potential of this direction. At the same time, it indicates the uncertainties and unknowns that need to be addressed in order to make AGN another established avenue to look for BSM physics in the Sky.

As the research activity on the AGN-BSM interplay is very recent, in Section~\ref{sec:AGN_BSM} we provide an executive summary of all such efforts that we know of, starting with DM and proceeding later with other BSM scenarios.
We then devote Section~\ref{sec:state_of_field} to the uncertainties and unknowns of this endeavor that have to do with our (so far) incomplete knowledge of AGN (Sec.~\ref{sec:AGN}) and of the distribution of DM around them (Sec.~\ref{sec:spikes}).
Finally, in Section~\ref{sec:questions} we identify some key open questions whose future answers, we believe, could help establish AGN as systems to learn solid BSM lessons.

This white paper is the result of a focus week at the Institute for Fundamental Physics of the Universe in Trieste~\cite{AGN_Dark_Sector_Laboratories_2026}, which brought together a community of experts in AGN observations and multi-messenger modeling, in supermassive BHs and DM around them, as well as in BSM phenomenology.

\section{BSM with AGN: an executive summary}
\label{sec:AGN_BSM}

\subsection{AGN for DM}

AGN are natural laboratories to understand the microphysical nature of DM, because DM in these environments is expected to be extremely dense and form so-called DM ``spikes''~\cite{Gondolo:1999ef} around the central supermassive BH fueling the AGN.
The highly non-thermal particles accelerated at AGN can therefore scatter off the ambient DM population, effectively turning these systems into
natural beam--dump experiments operating at energies far beyond those achievable on Earth -- and, in the case of blazars, we happen to lie right along the beamline! 

Indeed, the possible presence of DM spikes around AGN has recently been used to probe scattering interactions between DM particles and SM particles in AGN jets or coronae, see Fig.~\ref{fig:AGN_DM} for a sketch. Pioneering literature about AGN as probes of DM considered the possibility of testing GeV--TeV-scale DM scattering off cosmic-ray electrons and protons in AGN through electromagnetic signatures ~\cite{Bloom:1997vm,Gorchtein:2010xa,Huang:2011dg,Chang:2012sk,Profumo:2013jeb}. These exercises were also recently revisited, pointing out additional polarization signatures in \cite{Cermeno:2022rni}.

\begin{centering}
\begin{figure}
\includegraphics[width=0.45\textwidth,valign=c]{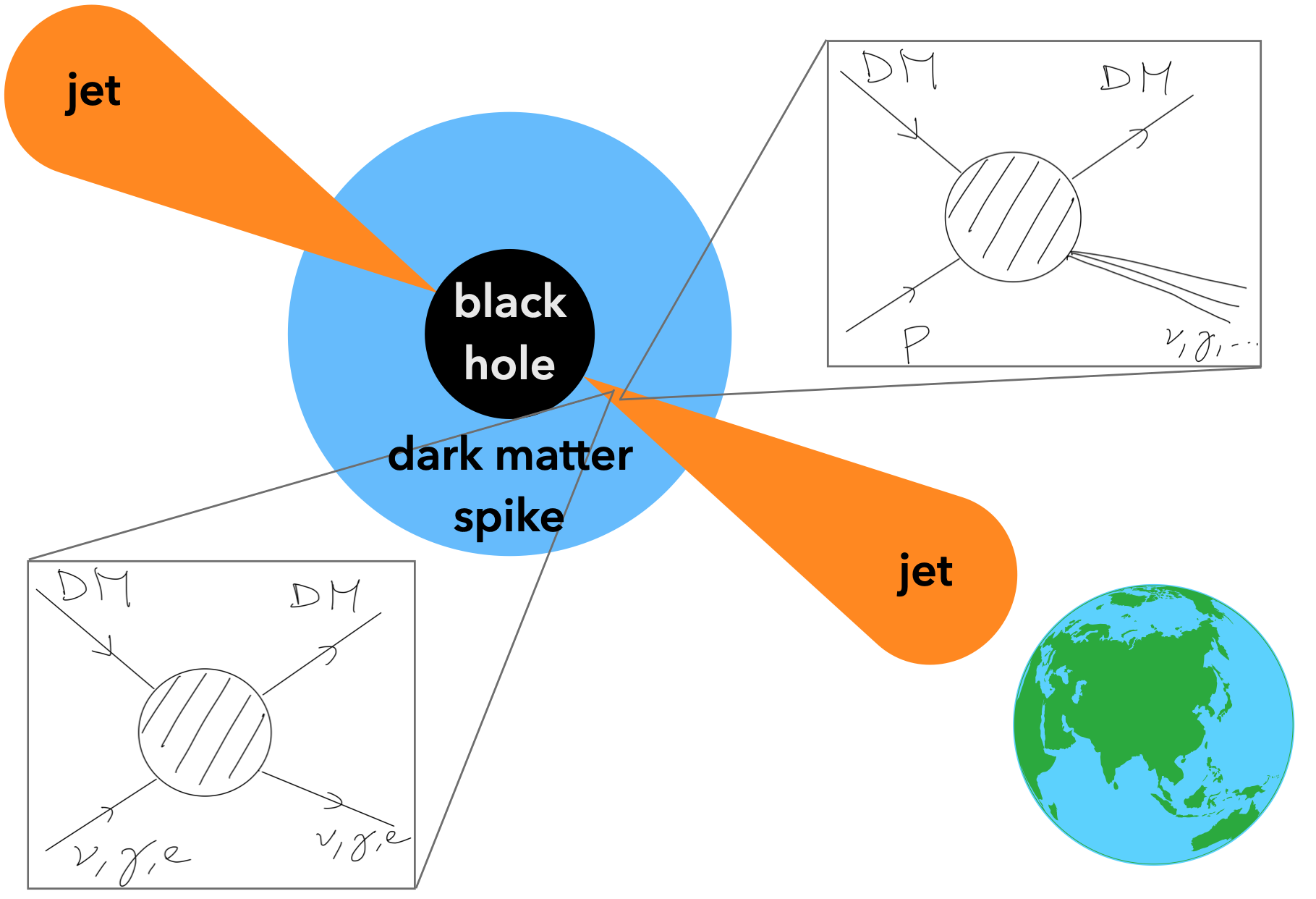}
\hfill
\includegraphics[width=0.53\textwidth,valign=c]{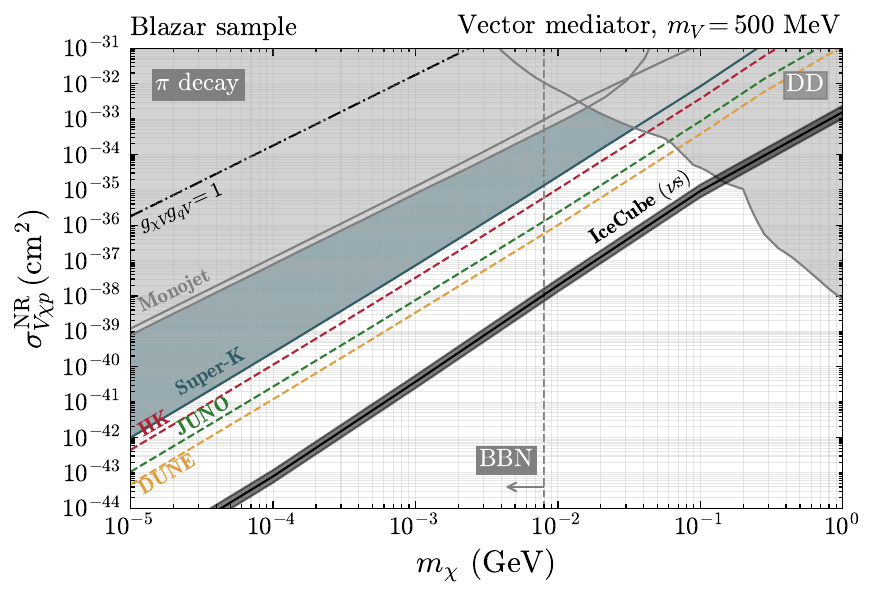}
\caption{\small  Left: sketch of dark matter density spike around the BH powering a blazar and of DM inelastic (top) and elastic (bottom) processes giving rise to multimessenger (neutrinos, gammas and boosted DM) signals on Earth. Right: DM-proton cross section $\sigma^{\rm NR}_{V\chi p}$ versus  mass $m_\chi$ of fermion DM, interacting with SM light quarks via a vector mediator $V$, where the gray shaded areas are excluded by pion decays (monojet at colliders give a comparable limit) and by DM direct detection. The blue shaded area and the dashed color lines represent, respectively, the limit and sensitivites of the indicated neutrino detectors to the  flux of DM as boosted by the sample of 324 blazar modeled in~\cite{Rodrigues:2023vbv}. The black `IceCube ($\nu$s)' line indicates the parameters for which the neutrinos produced by DM-proton scatterings, around the same blazars, saturate the IceCube observed diffuse flux. For DM mass below 150~MeV, the highest-energy diffuse neutrinos could be the first sign of a DM-SM interaction, compatibly with existing DM limits. Figure from~\cite{DeMarchi:2025uoo}, to which we refer for more details.
Other promising BSM probes that rely on AGN exist, that are not captured by the sketch on the left nor by the concrete example on the right, see text.\label{fig:AGN_DM}}
\end{figure}
\end{centering}

In recent years, it has been recognized that collisions of protons in blazar jets with a surrounding DM spike can %also
induce a substantial flux of boosted DM on Earth, that can be searched with neutrino and DM direct detection experiments~\cite{Wang:2021jic, Granelli:2022ysi}. Crucially, as in the case of galactic cosmic-ray-upscattered DM \cite{Bringmann:2018cvk, Ema:2018bih}, this process is particularly relevant for sub-GeV DM, as the corresponding non-accelerated halo component %sub-GeV DM 
lacks sufficient kinetic energy to induce nuclear recoils within the region of interest of such experiments, see e.g.~\cite{Essig:2022dfa}.
AGN-boosted DM signatures have since then been explored for various particle DM models~\cite{Bhowmick:2022zkj,Jeesun:2025gzt,Wang:2025ztb,DeMarchi:2025uoo}, including inelastic DM scenarios~\cite{Gustafson:2025dff}, from multiple blazars \cite{DeMarchi:2025uoo,Gago:2026ocx} and different AGN sources~\cite{Xia:2024ryt,Gustafson:2025dff}, 
considering contributions from deep inelastic proton-DM scatterings~\cite{Wang:2025ztb, DeMarchi:2025uoo,Gustafson:2025dff} and resonant ones~\cite{Gago:2026ocx}, as well as including detector response modelling and the impact of the host-galaxy DM profile~\cite{Barillier:2025xct}. 
AGN-boosted DM signals are mostly robust against caveats like spike depletion and time delays~\cite{DeMarchi:2025uoo}. Their detection would provide a smoking-gun signature of DM around these objects, making this a particularly compelling avenue for AGN-DM searches.

It was also pointed out that the high-energy neutrino and gamma-ray fluxes observed from certain AGN could also get sizably attenuated if a DM spike is present at these sources \cite{Cline:2022qld, Ferrer:2022kei,Cline:2023tkp}. This direction was further developed in several subsequent works, which extended the bounds on dark matter-neutrino scatterings to other high-energy neutrino sources ~\cite{Fujiwara:2023lsv}, accounted for time delay signatures~\cite{Fujiwara:2024qos}, considered different particle dark matter models, e.g.~a $B-L$ gauge boson~\cite{Cline:2026stf,Tseng:2024akh,Farzan:2026wek}, studied inelastic neutrino-DM scatterings~\cite{Chiang:2026dmc}, revisited the 2014--2015 neutrino outburst of TXS~0506+056~\cite{Zapata:2025huq}, explored DM-photon scatterings in NGC~1068~\cite{Herrera:2025gpm}, exploited the attenuation effects on ultra-high-energy neutrinos ~\cite{Bertolez-Martinez:2025trs,Mondol:2025uuw,Esteban:2025wbv,Dixit:2026zkv, Esteban:2026vzh}, and explored modifications of the flavor composition of AGN neutrinos~\cite{Abbaslu:2026row,Tseng:2026evn,Dev:2026kqu}.

It has subsequently been realized that elastic scatterings of DM with non-thermal particles in these sources also lead to an additional cooling mechanism~\cite{Herrera:2023nww}. The DM-induced cooling has also been studied for different particle DM models, including inelastic DM~\cite{Gustafson:2024aom,Mishra:2025juk} and deep-inelastic scatterings~\cite{Li:2025zwg,Chun:2026muh}. Within concrete cosmic-ray acceleration scenarios~\cite{Kantzas:2025huu,Meighen-Berger:2025hrq}, its observable consequences include modifications of the spectral energy distribution~\cite{Kantzas:2025huu} and of the spectral index of radio emission at different radii~\cite{Hussein:2025llu}. Inelastic collisions between protons in AGN jets and DM around them can break the proton apart and lead to additional high-energy neutrino~\cite{DeMarchi:2024riu, DeMarchi:2025xag} and gamma-ray~\cite{Chang:2012sk, Wang:2025ztb} emission. 

Some of these works have also interpreted AGN multi-messenger observations as arising from dark matter. Intriguingly, it has been proposed that the high-energy neutrinos arising from cosmic ray-DM interactions could be the origin of the neutrinos observed from TXS 0506+056~\cite{DeMarchi:2024riu}, a challenging observation to explain via $p$-$\gamma$ processes in single-zone models of blazar jets~\cite{Murase:2018iyl,Rodrigues:2023vbv,Gao:2018mnu,Cerruti:2018tmc,Keivani:2018rnh}, and even source part of the diffuse astrophysical neutrino flux measured by IceCube~\cite{DeMarchi:2025xag} when one extends this mechanism to a population of blazars, see Fig.~\ref{fig:AGN_DM} for an example in a concrete DM model. Complementary to neutrinos, gamma-rays have been shown to be a potentially more sensitive probe of proton-DM interactions~\cite{Wang:2025ztb}, but a conclusive statement requires a detailed study of their cascades within the source and a fit of the entire electromagnetic spectrum, including this additional contribution. In \cite{Herrera:2025gpm}, it has been discussed that the deficit of gamma-rays from NGC 1068 relative to the observed high energy neutrino fluxes may be induced by DM-photon scatterings. In \cite{Akita:2025dhg}, it was discussed that the neutrino emission from NGC 1068 may be due to DM annihilations in the spike. These works strengthen the idea that accounting for a DM component in AGN may help address some of the unresolved features of AGN and high-energy neutrino observations.

While most works relating AGN to DM have focused on probing DM-SM interactions in a model-independent manner, some (e.g.~\cite{Cline:2023tkp,Herrera:2023nww,DeMarchi:2024riu,DeMarchi:2025uoo}) have demonstrated that these probes are really testing new grounds of light DM, i.e.~they are testing DM models that are allowed by all its other existing probes (like those at colliders, underground detectors and telescopes) as displayed in Fig.~\ref{fig:AGN_DM}.
Recent multi-messenger AGN data could thus potentially shed more light on light dark sectors, across different particle channels and energy regimes.

\subsection{AGN for other BSM physics}
DM is not the only BSM scenario testable with AGN environments and multi-messenger data. Other possibilities include oscillations of gamma-rays into axion-like particles at the sources~\cite{Ayad:2019hrj,Davies:2022wvj,DeMarchi:2026pak} and during propagation~\cite{Chen:2017mjf,Malyshev:2025iis,Grimaud:2026hpu}, searching for modulations~\cite{Li:2020pcn,Davies:2022wvj,Malyshev:2025iis,DeMarchi:2026pak}, polarization signatures~\cite{Yao:2022col}, or shifts in the expected normalization of the astrophysical gamma-ray fluxes from these sources~\cite{Ayad:2019hrj}. In some cases, the astrophysical expectation can be anchored with high-energy neutrinos through a multi-messenger fit~\cite{Dekker:2025vcg}. Further, the production of dark sector particles, gravitons, BHs and model-independent bounds in $pp/p\gamma$ collisions within the AGN has also been recently discussed \cite{Dev:2025czz, Ettengruber:2025kat, Arguelles:2026rjr}.

Rapidly spinning supermassive BHs in AGN can probe ultralight bosons through black-hole superradiance, with sensitivity to masses $m_{\phi}\sim10^{-21}-10^{-17}\mathrm{eV}$ ~\cite{Arvanitaki:2014wva,Unal:2020jiy}. Observable signatures include depleted regions in the SMBH mass--spin plane and nearly monochromatic gravitational waves from boson-cloud annihilations or transitions~\cite{Cardoso:2018tly,Baryakhtar:2017ngi,Brito:2017wnc}. If the bosons couple to Standard Model fields, the cloud may also produce polarization oscillations, plasma emission, or broader modifications of AGN spectra and accretion~\cite{Chen:2019fsq,Siemonsen:2022ivj,Sarmah:2024nst}.

\section{The state of the field(s)}
\label{sec:state_of_field}

\subsection{Active Galactic Nuclei}
\label{sec:AGN}

The detection by IceCube of high-energy neutrinos towards at least two AGN sources, NGC 1068 and TXS 0506+056~\cite{IceCube:2022der, IceCube:2018cha}, has opened the era of multi-messenger studies of AGN. This has complemented decades of observations of AGN emission across the electromagnetic spectrum, from radio, optical, X-rays to gamma-rays, see, e.g., Refs.~\cite{IceCube:2018dnn,MAGIC:2022gyf,Fermi-LAT:2022byn,Garc_a_Burillo_2016,Impellizzeri_2019}.

The possibility of producing high-energy astrophysical neutrinos through cosmic-ray interactions was recognized more than five decades ago~\cite{BERESINSKY1969423,Berezinsky:1975zz}, with AGN subsequently identified as promising neutrino sources~\cite{Eichler:1979yy,Berezinsky:1981sdl,Stecker:1991vm}; see, e.g., Ref.~\cite{Murase:2022feu} and references therein for a modern overview. The detections of NGC~1068 and TXS~0506+056 now provide direct evidence that AGN accelerate protons well beyond the TeV scale, and possibly up to the EeV scale~\cite{Rodrigues:2025cpm,KM3NeT:2025npi}. These cosmic rays undergo $pp$ and $p\gamma$ interactions at proton energies far above those reached in other astrophysical accelerators such as supernova remnants; for $pp$ collisions the corresponding center-of-mass energies exceed those of terrestrial colliders, providing a novel opportunity to test physics beyond the Standard Model.

\subsubsection{Nomenclature}
AGN are broadly classified according to their observational properties. First, the presence or absence of relativistic, large-scale jets, perpendicular to the black hole accretion disc, determines whether they are radio-loud or radio-quiet objects, owing to the synchrotron radio emission from electrons in the jets. In the following, we adopt the nomenclature \textbf{jet-loud}/\textbf{jet-quiet} to identify these two AGN classes.
 
In addition to this dichotomy, AGN appear observationally different due to their orientation to the observer. The central region is surrounded by a dust torus that allows a view of the BH and its accretion disc only when the system is observed face-on. This leads to further classifications:
\begin{itemize}
\item[$\diamond$] Among jet-quiet AGN, \textbf{Seyfert} galaxies are the most common class. They are defined by moderate bolometric (i.e. total) luminosities, roughly  $10^{42}-10^{45}$ erg/s (as opposed, e.g., to the $10^{45}-10^{48}$ erg/s of quasars \cite{Urry:1995mg, Padovani:2017zpf}).
They are  further classified as Seyfert-1 (unobscured, face-on view) and Seyfert-2 (obscured, edge-on view).
\item[$\diamond$] Jet-loud AGN also have an orientation classification: when the jets are observed edge-on they are called radio-galaxies, while when one of the jets is directed very nearly towards the observer the source is called a \textbf{blazar}. Blazars are characterised by a spectral energy distribution dominated by a non-thermal continuum that covers the whole electromagnetic spectrum, from the radio up to gamma-rays. Gamma-ray blazars are classified primarily into BL Lacertae objects (BL Lacs) and flat-spectrum radio quasars (FSRQs), according to whether their optical spectra exhibit weak or absent emission lines, or strong and broad emission lines, respectively. This distinction is generally interpreted as reflecting different accretion regimes and external radiation fields, with the gamma-ray emission being dominated by synchrotron self-Compton processes in BL Lacs and by external Compton scattering in FSRQs.
\end{itemize}

\subsubsection{AGN observations in gamma-rays and neutrinos}
\label{sec:AGN_observations}

Blazars constitute the brightest and most abundant class of sources in the extragalactic gamma-ray sky, primarily because their relativistic jets are closely aligned with the observer’s line of sight. The 16-year catalogue compiled from data from the Large Area Telescope (LAT) on board of the Fermi Gamma-ray Space Telescope contains a list of 7,220 sources, 4,173 of which are classified as blazars or blazar candidates, providing a rich data set for investigating the physical properties and diversity of their subclasses~\cite{Ballet:2026kpq}. The planned CTAO extragalactic survey will cover approximately one quarter of the sky and extend spectral and temporal measurements of blazars to multi-TeV energies~\cite{CTAConsortium:2017dvg}.

A key role is also played by the analysis of the isotropic gamma-ray background (IGRB) which comprises the emission from unresolved extragalactic sources. Predictions for such background depend on the cosmological evolution encoded in the luminosity functions of misaligned AGN, BL Lacs and FSRQs~\cite{DiMauro:2013xta,Ajello:2011zi,Ajello:2013lka,Qu:2019zln,Blanco:2021icw}. BL Lacs are particularly important since models suggest that their hard spectra might extend to multi-TeV energies. Absorption of this emission on the extragalactic background light initiates electromagnetic cascades that redistribute its power into the GeV range. Early population calculations already indicated that blazar cascades could substantially alter the extragalactic gamma-ray background, depending on the luminosity function and the extragalactic magnetic field~\cite{Murase:2012df,Yan:2012xs}. More recent conservative models of resolved blazars have identified a potential tension between this expected cascade emission and the measured IGRB, disfavoring scenarios in which the multi-TeV power of blazars is efficiently transferred into GeV $\gamma$-rays~\cite{Blanco:2023kfa}. Model uncertainties propagate directly into constraints on additional components, such as a diffuse gamma-ray flux from DM pair annihilations or decays~\cite{Fermi-LAT:2015qzw,Murase:2012xs,Murase:2015gea,Blanco:2018esa,Das:2023wtk}.

\medskip

Gamma-ray observations have been recently complemented by high-energy neutrino surveys. The IceCube Neutrino Observatory~\cite{Aartsen_2017}, built deep in the Antarctic Ice, has been looking for neutrino signals from AGN since its construction. The most significant association between neutrinos and these sources comes from the Seyfert galaxy NGC 1068,  with a significance of $4.2\sigma$~\cite{IceCube:2022der}. The detection of NGC 1068 over the years is consistent with persistent neutrino emission from this galaxy. The updated 13.1-year analysis finds a soft spectrum, with the 95\% central energy interval of events contributing to the NGC 1068 excess spanning $0.2$--$20.6\,\mathrm{TeV}$~\cite{Abbasi_2026}. Additional Seyfert candidates, including NGC 4151 and CGCG 420-015, show neutrino excesses with significances of about $3\sigma$, motivating searches for a collective contribution from AGN to the diffuse astrophysical neutrino flux~\cite{IceCube:2024ayt,IceCube:2024dou}. 
Recent population searches report a global significance of $3.3\sigma$ for a collective excess from 11 of 47 northern X-ray-bright AGN, identified using a binomial test, and a significance of $3.0\sigma$ from a model-weighted stacking analysis of 13 southern Seyfert galaxies~\cite{Abbasi_2026,Abbasi_2026_2}.

The other promising candidate of a neutrino source class is blazars. After the multimessenger detection of the blazar TXS$\,0506+056$~\cite{IceCube:2018dnn}, in which a high-energy neutrino was found in spatial and temporal coincidence with a gamma-ray flare from the blazar, many studies have tried to identify neutrino emission from these objects.  Searches have targeted both individual sources, for example, time-dependent emission from TXS$\,0506+056$~\cite{doi:10.1126/science.aat2890}, as well as collective excesses in blazar catalogs~\cite{IceCube:2025tmc,Abbasi_2023}. Although the archival TXS$\,0506+056$ search reported $3.5\sigma$ evidence for neutrino emission, catalog-level searches found no statistically significant excess across the population~\cite{IceCube:2018cha,IceCube:2025tmc,Abbasi_2023}. Assuming flavor equipartition at Earth and a single power-law spectrum with index $-2.5$,
blazars in the second catalogue of AGN from Fermi contribute no more than 27\% of the observed astrophysical neutrino flux between about $10\,\mathrm{TeV}$ and $2\,\mathrm{PeV}$ at 90\% C.L.~\cite{IceCube:2016qvd}. The gamma-ray--neutrino connection strongly constrains gamma-ray transparent sources of the 10--100~TeV neutrino flux~\cite{Murase:2013rfa,Murase:2015xka,Bechtol:2015uqb}, since blazars dominate the extragalactic gamma-ray background above $\sim 50$~GeV~\cite{Fermi-LAT:2015otn,Lisanti:2016jub,Ajello:2015mfa} yet contribute only a subdominant fraction of the neutrino flux~\cite{IceCube:2016qvd,Hooper:2018wyk,Rodrigues:2023vbv}. This favors a population of gamma-ray--opaque sources~\cite{Murase:2015xka,Capanema:2020rjj,Inoue:2019yfs}, as now supported by NGC~1068~\cite{IceCube:2022der} and by the population analysis of X-ray-bright AGN in the northern sky~\cite{Abbasi_2026}.

Constraints on individual ultra-high-energy neutrino sources remain comparatively weak because expected event counts are low and current experiments have limited effective areas. Next-generation experiments, including RNO-Greenland, which is under construction, and the proposed IceCube-Gen2, are designed to improve sensitivity to neutrinos at these energies~\cite{RNO-G:2020rmc,IceCube-Gen2:2020qha}.

\begin{figure}[t]
\centering
\includegraphics[width=0.45\textwidth,valign=c]{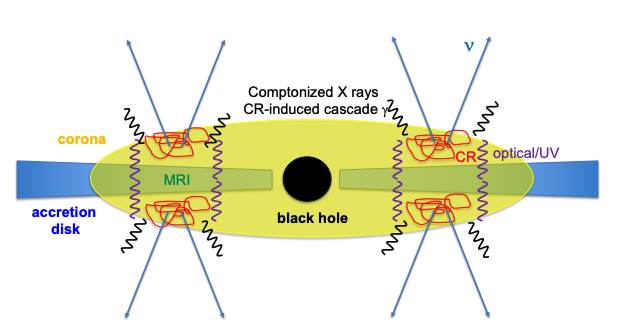}
\raggedright
\includegraphics[width=0.46\textwidth,valign=c]{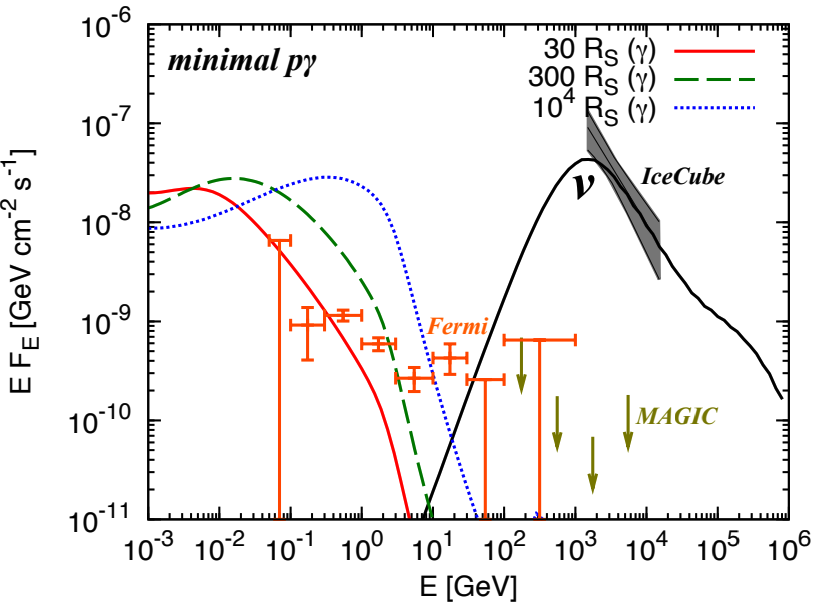}
\\[0.4em]
\includegraphics[width=0.46\textwidth,valign=c]{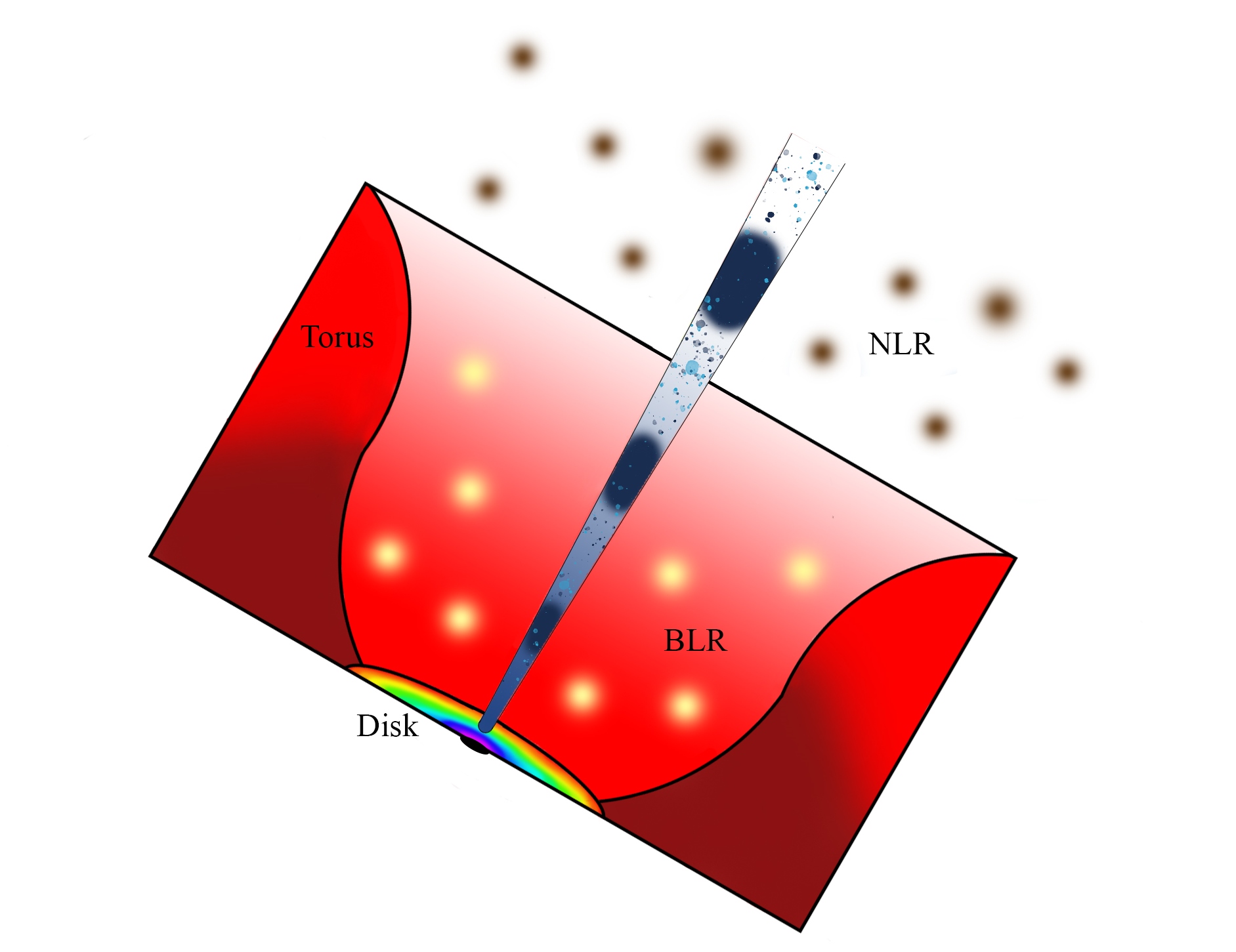}
\hfill
\includegraphics[width=0.53\textwidth,valign=c]{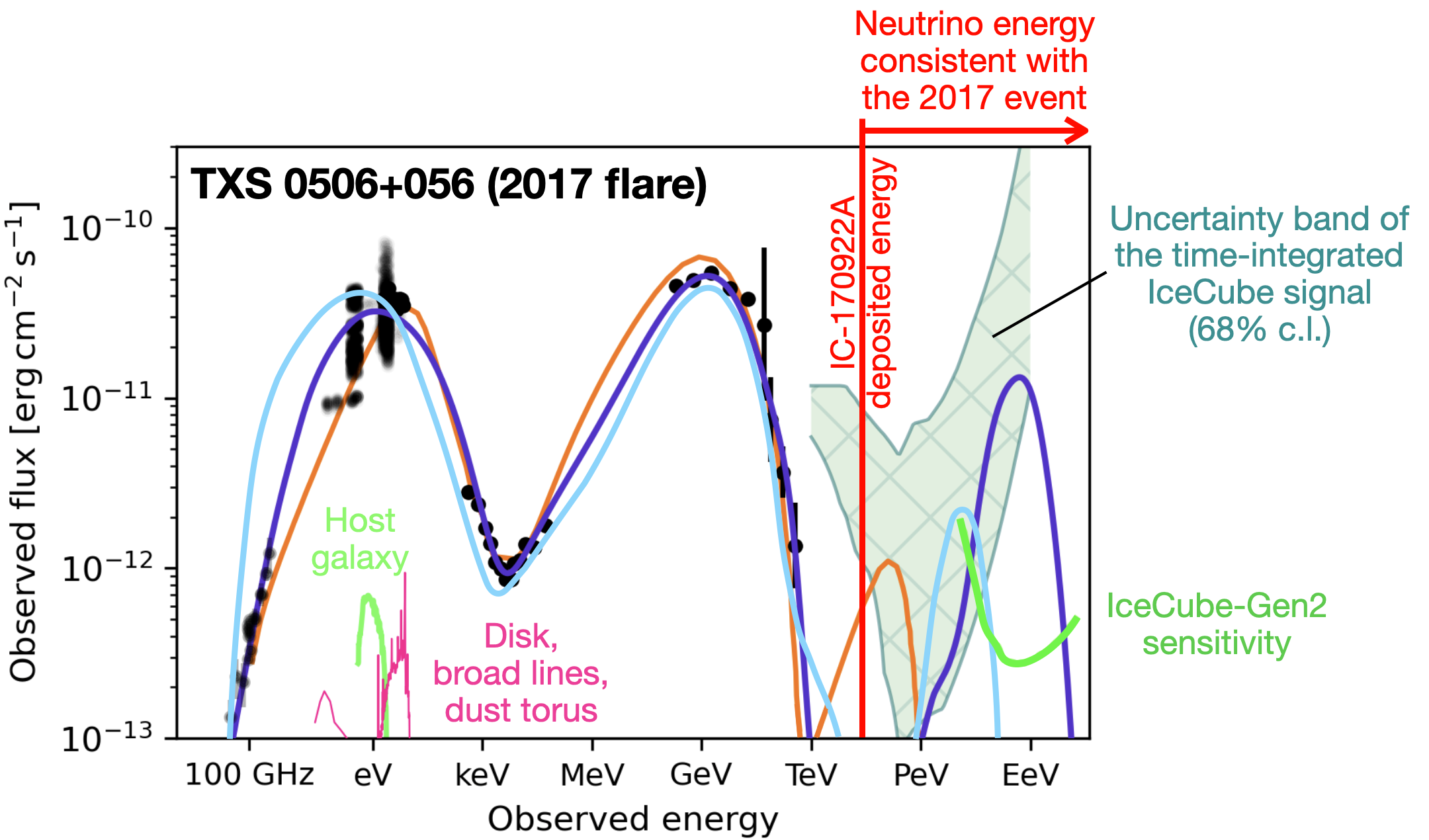}
\caption{\small
Top: in jet-quiet AGN like NGC~1068, protons accelerated in the hot corona above the accretion disk produce neutrinos and gamma-rays. Neutrinos escape the AGN, gamma-rays are absorbed and reprocessed to lower energies if the emission comes from within a few tens of Schwarzschild radii of the black hole~\cite{Murase:2019vdl}.
This explains why IceCube sees a high neutrino flux (black solid line) compatibly with Fermi gamma-ray measurements and MAGIC upper limits (red solid line; dashed and dotted lines are the gamma-ray prediction in case of larger emission regions); a large MeV gamma-ray flux is expected within this picture~\cite{Murase:2019vdl,Ajello:2023hkh}.
Bottom: in jet-loud AGN, particles are accelerated in blobs moving along the relativistic jet, bathed in photons from the disk, broad-line region (BLR), dusty torus and narrow-line region (NLR)~\cite{Cerruti:2020lfj}. Three lepto-hadronic models of the 2017 flare of the blazar TXS~0506+056 (coloured curves) reproduce the electromagnetic data (black) well, yet predict neutrino fluxes peaking at different energies and with different amplitude~\cite{Rodrigues:2026kpf}.}
\label{fig:AGN_corona}
\end{figure}

\medskip

AGN are also observed across the electromagnetic spectrum, from radio frequencies through the infrared, optical/UV, and X-ray bands to gamma rays~\cite{Padovani:2017zpf}. When combined with neutrino data, these observations can constrain the emission regions, as shown for the flaring blazar TXS~0506+056~\cite{Keivani:2018rnh,Gao:2018mnu,Cerruti:2018tmc,Rodrigues:2018tku,Petropoulou:2019zqp} and for the corona of NGC~1068~\cite{Inoue:2019yfs,Murase:2022dog,Eichmann:2022lxh,Fiorillo:2023dts,Mbarek:2023yeq}, help discriminate between hadronuclear ($pp$) and photohadronic ($p\gamma$) neutrino-production scenarios~\cite{Eichmann:2022lxh,Das:2024vug}, and perhaps constrain cosmic-ray acceleration mechanisms~\cite{Fiorillo:2023dts,Das:2026rwt}. In particular, UV and X-ray measurements constrain the radiation fields relevant for photohadronic interactions, while X-ray observations can bound the electromagnetic cascades accompanying neutrino production. Although exploiting these electromagnetic observations as probes of BSM physics would be worthwhile, such studies are lacking.

\subsubsection{AGN multi-messenger astrophysical models}

The energy spectra, luminosities and spatial location of high-energy particles around AGN, as well as the medium they interact with, are crucial inputs to compute any BSM signal related to AGN.
Their modeling incorporates astrophysical knowledge about these systems and aims at reproducing the multi-messenger observations summarised in Sec.~\ref{sec:AGN_observations}.
We now summarise the current status of these models, first for jet-quiet and then for jet-loud AGN.

\paragraph{Jet-quiet AGN}
Following the IceCube evidence for TeV neutrino emission from NGC 1068, jet-quiet AGN, particularly Seyfert galaxies, have emerged as promising sources of high-energy neutrinos. Early models invoked proton acceleration at accretion shocks near the black hole~\cite{Stecker:1991vm}, whereas more recent studies have explored particle acceleration in the coronae surrounding the accretion disk~\cite{Inoue:2019yfs,Murase:2019vdl,Kheirandish:2021wkm,Fiorillo:2023dts}. 
In these systems, the compact, magnetized corona surrounding the accretion disk provides a natural site for nonthermal proton acceleration through magnetic reconnections and/or plasma turbulence. Accelerated protons interact with the dense coronal radiation and matter fields through photomeson and hadronuclear processes, producing neutrinos, gamma rays, and secondary electron–positron pairs. Because the corona is highly compact and opaque to high-energy photons, the gamma rays accompanying neutrino production are absorbed and reprocessed through electromagnetic cascades, whereas neutrinos escape freely. AGN coronae can therefore operate as “hidden” cosmic-ray accelerators, potentially producing substantial neutrino emission without violating the observed GeV–TeV gamma-ray background~\cite{Murase:2019vdl,Murase:2026hrz}, see Fig.~\ref{fig:AGN_corona} (top). Alternatively to the corona model, the neutrino emission region has been proposed to be linked to the powerful winds launched by the accretion disk. In this scenario as well, the emitting region is opaque to gamma-ray photons due to pair-production, and the electromagnetic emission is reprocessed to lower energies via pair cascades \cite{SInoue22, Peretti23}. 

NGC 1068 provides the most important observational test of this scenario. While IceCube reported an excess of TeV neutrinos from its direction, its observed GeV gamma-ray flux is substantially lower than the electromagnetic power expected to accompany optically thin hadronic neutrino production (see Fig.~\ref{fig:AGN_corona}, top right). This requires a compact neutrino-emitting region deeply embedded in the AGN nucleus rather than the spatially extended starburst region or a gamma-ray-transparent outflow. Detailed coronal models have reproduced the measured neutrino spectrum if a significant fraction of the magnetic or accretion power is transferred to nonthermal protons. Recent post-detection studies using neutrino and gamma ray data find that viable coronal scenarios favor strong magnetic fields and compact emission regions, with substantial power channeled into nonthermal protons~\cite{Murase:2022dog,Blanco:2023dfp,Das:2024vug,Das:2026rwt}.  Population models predict that NGC 1068-like jet-quiet AGN could contribute substantially to the diffuse neutrino intensity around 10~TeV~\cite{Murase:2019vdl,Murase:2026hrz}.

\paragraph{Jet-loud AGN} 
From a theoretical standpoint, the gamma-ray emission from blazars can originate in a population of non-thermal electrons, via inverse Compton scattering of the ambient photon fields, or non-thermal protons (or higher-mass isotopes), via synchrotron emission or hadronic processes such as pion production~\cite{Boettcher:2013wxa,Cerruti:2020lfj}. If protons undergo hadronic interactions, neutrinos are co-produced via the decay chain of the charged pions. The kinematics of the reaction dictates that the neutrinos carry about 5\% of the energy of the parent proton, while the gamma-rays produced in the decay of the neutral pion carry about 10\% of the proton's energy. Crucially, while neutrinos escape the source, gamma rays can undergo cascading electromagnetic interactions that redistribute their energy to lower frequencies. From this simple picture, it is clear that \textit{(i)} the multi-wavelength signatures of neutrino-emitting blazars are not limited to simultaneous gamma-ray emission, but may be far more complex both in the time and frequency domain; and \textit{(ii)} not all bright blazars are necessarily strong neutrino emitters~\cite{Murase:2014foa}, as their gamma-ray emission can be driven purely by electrons~\cite{Boettcher:2013wxa}. Together, these two problems represent current obstacles faced by the theory community in developing predictive models for interpreting the neutrino observations described in the previous subsection.

Problem \textit{(i)} has to do chiefly with the understanding of the source physics. Because of the non-linear nature of the photohadronic interactions described previously, the system has to be computed numerically. To this end, the theory community has developed several numerical leptohadronic codes over the past decade. In an attempt to grasp the systematic differences between these codes, a recent work has compared them head-to-head~\cite{Cerruti:2024lmj}. Under the same conditions, the authors performed simulations of blazar jets in a simple, single-zone model, in which the dominant non-thermal emission region is approximated as a single, homogeneous, spherical plasma blob moving relativistically along the jet axis, and compared the multi-wavelength steady-state spectrum predicted by each code. For scenarios involving only electron emission, different codes yielded frequency-dependent multi-wavelength fluxes, or spectral energy distributions, differing by no more than 10\%. For realistic proton synchrotron and leptohadronic models, the codes agreed on the spectral shapes but showed a maximum spread of $\pm40\%$ in the overall normalization. This should be understood as a conservative systematic uncertainty when comparing simulations with observations~\cite{Cerruti:2024lmj}.

In contrast, solving problem \textit{(ii)} would require a predictive framework to confidently identify multi-wavelength signatures of neutrino emission from blazars. This lies beyond the reach of current models, owing to (a) the complex nature and wide variety of multi-wavelength leptohadronic signatures, as we exemplified above with the frequency-smearing of the gamma-ray emission via electromagnetic cascades; (b) the limited spatial resolution of observational data from blazars, which are distant sources and highly beamed toward the observer; and (c) strong degeneracies intrinsic to state-of-the-art source models~\cite{Rodrigues:2026kpf}, as illustrated for TXS~0506+056 in Fig.~\ref{fig:AGN_corona} (bottom right). On the observational end, progress may require, among other possibilities, the inclusion of more fine-grained time-domain data, high-resolution radio imaging with the very-large-baseline interferometry (VLBI) technique, or multi-wavelength polarization data~\cite{Liodakis:2021els}. The lack of gamma-ray instruments sensitive in the MeV range has also been noted as a major gap in constraining models of blazars and AGN at large~\cite{Reimer:2018vvw,Rodrigues:2018tku,Murase:2019vdl,Zhang:2019htg,Biteau:2020prb,Murase:2023ccp}. While AMEGO-X and e-ASTROGAM remain proposed, COSI is scheduled for launch in 2027 and will provide coverage from $0.2$ to $5\,\mathrm{MeV}$~\cite{Caputo:2022xpx,e-ASTROGAM:2017pxr,Tomsick:2023aue}. Simultaneously, from the theoretical side, it is necessary to develop more complex models that can hopefully predict these different features in an increasingly self-consistent fashion~\cite{Zhang:2019dob,Zacharias:2022pea,Rodrigues:2025cpm}.

\subsection{Dark matter around supermassive black holes}
\label{sec:spikes}
In addition to being efficient sources of very energetic particles, AGN are expected to reside in environments with extremely high DM densities. An adiabatic growth of the central supermassive BH induces a sharp steepening of the DM density profile in the region of gravitational influence of the BH, leading to the formation of a so-called dark matter 
``spike''~\cite{Gondolo:1999ef}. In the following, we briefly review the theoretical modeling of these spikes and the observational prospects to probe them in, respectively, Sec.~\ref{sec:spikes_models} and~\ref{sec:measurements_spikes}.
Estimates of BH masses are a key ingredient for inferring spike densities; in Sec.~\ref{sec:measurements_MBH} we summarize relevant black hole mass measurements.

\subsubsection{Modeling dark matter spikes around black holes}
\label{sec:spikes_models}

If the growth of a SMBH at the center of a collisionless system proceeds adiabatically, namely on a timescale much longer than the orbital time of the surrounding particles, the resulting contraction of the system can be followed semi-analytically exploiting adiabatic invariants~\cite{Quinlan:1994ed,Gondolo:1999ef}. As the gravitational potential deepens, individual particle orbits contract, producing an enhancement of the central density. In particular, under the assumptions of spherical symmetry and of an originally isotropic phase-space distribution function, and considering an initial DM halo with power-law radial density profile $\rho\propto r^{-\gamma}$ ($\gamma>0$), the framework leads to a set of predictions for the final DM spike density profile: 

\textit{(i)} The power-law index steepens to $\gamma_\mathrm{sp}=(9-2\gamma)/(4-\gamma)$. E.g., assuming as initial DM configuration the inner slope $\gamma \simeq 1$ of the Navarro–Frenk–White (NFW) profile~\cite{Navarro:1995iw}, a standard analytical model that describes results of cosmological N-body simulations of hierarchical cold DM clustering, one finds $\gamma_\mathrm{sp}=7/3$~\cite{Quinlan:1994ed,Gondolo:1999ef}. 

\textit{(ii)} This steepening extends from twice the Schwarzschild radius of the SMBH, below which no stable orbit is allowed~\cite{Sadeghian:2013laa}, up to a fraction ($\sim 0.2$) of the radius of gravitational influence of the black hole, defined as the radius within which the enclosed DM mass is twice that of the central BH~\cite{Merritt:2003}.

\textit{(iii)} The overall normalization of the spike density, tentatively $\rho \sim M_\mathrm{BH}^{2/3}\,c^{1/2}\,M_\mathrm{halo}^{1/9}$,
has a weak dependence on the halo mass $M_\mathrm{halo}$ and is governed predominantly by the BH mass and the concentration parameter $c$. Here, the assumption is that the inner spike smoothly connects to the outer DM density profile specified by a parametric form in terms of a mass normalization (the total mass $M_\mathrm{halo}$ within the virial radius $R_\text{vir}$) and a characteristic scale (the scale radius $r_s$, defined as some intermediate radius at which $d\ln\rho_\mathrm{halo}/d\ln r\big|_{r=r_s} = -2$; the concentration parameter is, in turn, defined as $c = R_\text{vir}/r_s$). This is the case, e.g., for the NFW profile; halos with larger $c$ are more ``centrally concentrated'', hence feature a larger initial population of DM particles within the BH radius of influence.     

Variants of this idealized picture, possibly better capturing features in the formation of DM spikes and their evolution, have been explored with semi-analytic models and/or numerical simulations in a number of works in the last two decades~\cite{Ullio:2001fb, Merritt:2002vj,Merritt:2003qk,Bertone:2005hw, Vasiliev:2008uz,Sadeghian:2013laa,Ferrer:2017xwm,Shapiro:2022prq,Bertone:2024wbn,Caiozzo:2025mye,Sharpe:2026nqq, Herrera:2026hal, Karydas:2026ctq,Sharpe:2026kfk}. The focus has been mostly on halos and BHs up to the low-mass tail of SMBHs, since this is relevant for indirect detection signals of DM pair annihilations in a putative DM spike around Sagittarius A* (Sgr A*), the $4\cdot 10^6 \, M_\odot$ BH at the Galactic Center, 
see, e.g., \cite{Gondolo:1999ef, Belikov:2013nca, Lacroix:2015lxa, Shelton:2015aqa}, as well as for the orbital evolution in Extreme Mass Ratio Inspiral (EMRI) events potentially detectable by the LISA gravitational wave telescope, see, e.g.,~\cite{Eda:2013gg,Eda:2014kra,Macedo:2013qea,Barausse:2014tra,Karydas:2024fcn,Kavanagh:2024lgq}. While rarely critically addressed in the context of AGN, we group the main uncertainties in modeling DM spikes into the following categories:
\begin{centering}
\begin{figure}
\includegraphics[width=0.49\textwidth,valign=c]{Regrowth.pdf}
\hfill
\includegraphics[width=0.49\textwidth,valign=c]{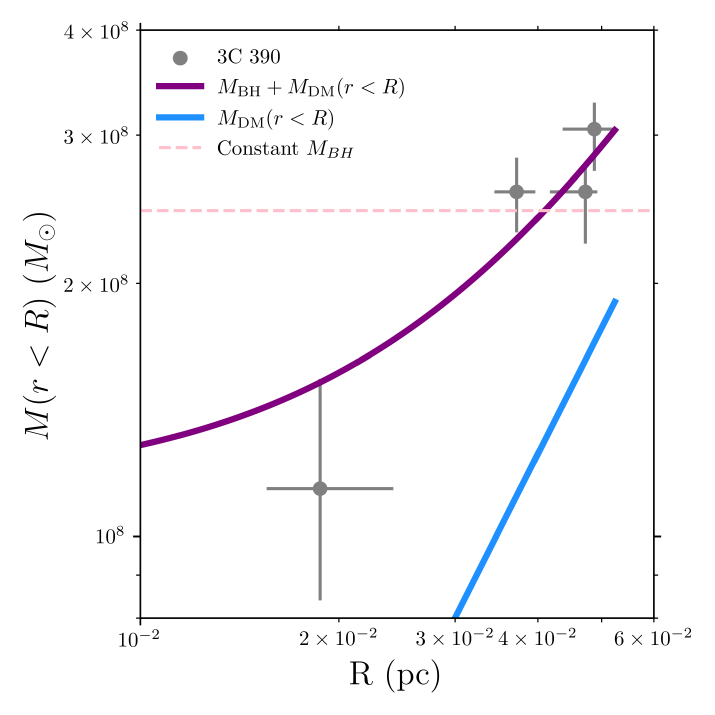}
\caption{\small Left: DM density around a BH versus distance from it. 
The plot shows the standard spike produced by the adiabatic growth of a central BH (dashed red), together with the ``mound'' resulting from the adiabatic growth of a supermassive star followed by its collapse into a $10^5 \, M_\odot$ BH (solid red). Also shown are the profiles obtained when the collapse is followed by a subsequent phase of adiabatic growth of the central BH, increasing its mass by a factor of 2 or 5.6~\cite{Caiozzo:2025mye}. Differences between spike and mound profiles are appreciable only close to the BH and tend to be further erased by a growth phase.  
Right: mass enclosed within a radius $R$ of the AGN 3C~390, measured with reverberation mapping (points); a DM spike (blue) would make it grow with radius (purple), unlike a black hole alone (dashed pink)~\cite{Sharma:2025ynw}.\label{fig:Spikes}}
\end{figure}
\end{centering}

\begin{enumerate}

\item Uncertainties from the mechanism of SMBH formation and growth. If the black hole in the DM halo forms off-center, or if its growth proceeds through non-adiabatic changes in the potential, the resulting profile can be significantly less steep and less dense than the standard adiabatic-spike formalism would suggest~\cite{Ullio:2001fb,Goerdt:2008}. Hybrid scenarios, which may be more realistic in case of AGN, can preserve most of the standard spike structure. For instance, considering a direct-collapse-black-hole seed and a subsequent adiabatic growth,  deviations are confined within a few tens of Schwarzschild radii and may be further erased in the final stage of growth~\cite{Bertone:2024wbn,Caiozzo:2025mye}, 
see Fig.~\ref{fig:Spikes},
with a very mitigated impact on BSM signals from DM-interactions with particles in the AGN. 

\item Depletion through interactions with stars and compact objects. Dynamical friction from celestial bodies can locally heat the DM particles and reshape the profile, scattering DM into larger orbits, which could reduce the density of DM in dense stellar environments or in the presence of inspiralling BHs~\cite{Bertone:2005hw,Merritt:2006mt,Kavanagh:2020cfn,Sharpe:2026nqq,Herrera:2026hal,Karydas:2026ctq,Sharpe:2026kfk}.

\item The merger history of the host. Equal-mass mergers of SMBHs are known to possibly deplete the spike \cite{Merritt:2002vj,Kavanagh:2018, Sharpe:2026nqq}. 
To our knowledge, studies of the impact of mergers on the spikes around a population of AGN have not been performed.
The physics case exposed in this white paper motivates them.
Existing results for SgrA*~\cite{Merritt:2002vj} suggests that the number of DM overdensities around AGN, that survive mergers, will depend on the fraction of SMBHs that ultimately merge at comparable masses.

\item Uncertainties from the DM particle properties. DM pair annihilation softens the innermost profile, replacing the naive cusp with a shallower ``weak cusp'' ($\rho \sim r^{-1/2}$ for $s$-wave, and even flatter for $p$-wave)~\cite{Vasiliev:2007,ShapiroShelton:2016}. DM self-interactions likewise produce a central overdensity, but with a steepness and normalization that depend on the scattering cross section and its velocity dependence~\cite{Shapiro:2018,ShapiroPaschalidis:2014}. Finally, mass or energy segregation mechanisms, analogous to those that take place in the formation of stellar cusps~\cite{2004ApJ...613L.109P} might be involved if DM and baryonic matter do couple other than gravitationally.
Some of these effects have been taken into account in the context of using AGN as BSM laboratories, e.g., proving in~\cite{DeMarchi:2025uoo} that they marginally affect the signals of boosted DM and of neutrinos from DM around blazars.

\end{enumerate}

\noindent
The modeling of some of the effects above is sometimes performed via numerical N-body simulations, a tool requiring some caution as it can introduce non-physical effects.
For example, discreteness effects in $N-$body simulations can in principle drive away a massive particle initially at rest in the centre of the system, an effect qualitatively similar to the so-called BH wandering in stellar systems \cite{2023A&A...673A...8D}. Moreover, the discreteness effects associated to the simulation resolution (i.e. the number of particles $N$ representing the DM distribution) might also partially hamper the dynamical friction on an inspiralling BH, thus leading to overestimate the infall times~\cite{2025A&A...700A.230D}.

Finally, studies of DM spikes have mostly been performed assuming isotropy and spherical symmetry, and their results may be affected by structures in the phase space distribution.
For example, there might be collisionless collective processes driving a central BH out from the centre of the spike or significantly reducing dynamical friction. The effect, usually known as ``dynamical buoyancy"~\cite{2021ApJ...912...43B,2022ApJ...926..215B}, occurs when a coupled underdensity-overdensity behaves as a massive quasi-particle and may kick out the central BH. The onset of this instability is most likely associated to the phase-space structure of the DM distribution \cite{2025arXiv251111804D}, and further studies are needed to assess the relevance of this effect in the context of AGN.

\subsubsection{Mass measurements of black holes}
\label{sec:measurements_MBH}

Most techniques for measuring BH masses are limited to systems in the local Universe. This is in part due to the fact that resolving the relevant dynamical scales becomes increasingly challenging with distance, and in part because an accurate distance is needed to convert angular scales on the sky to physical scales in the host galaxy.  
Stellar dynamical modeling and gas dynamical modeling (see, e.g., the review by \cite{Kormendy2013}) are limited to distances $D \lesssim 100$ \,Mpc. They probe the mass enclosed by the BH sphere of influence, 
    $r_{\rm infl}=GM_{\rm BH}/\sigma_*^2 \approx 11.2 (M_{\rm BH}/10^8 M_{\odot}) / (\sigma_*/200 {\rm km/s})^2$ pc,
where $M_{\rm BH}$ is the BH mass, $\sigma_*$ is the bulge stellar velocity dispersion, and $G$ is the gravitational constant. 
Masers can be resolved out to slightly larger distances ( $\lesssim 200$\,Mpc; \cite{Rosenthal2020}) and can probe the mass within $0.1-1$\,pc of the center, although systems with the appropriate geometry and physical conditions are extremely rare (see, e.g., the review by \cite{Lo2005}).

For BH mass measurements at cosmological distances, a leading technique to measure their mass is reverberation mapping of broad-lined AGN~\cite{Blandford1982,Peterson1993}. This probes the mass within light-days of the central BH ($0.001$--$0.1$~pc) and has been successfully applied out to $z<3$~\cite{Shen2024}. Trading spatial for temporal resolution, it measures time delays between flux variations arising from different components of the AGN, thereby probing physical scales deep within the BH gravitational potential. Although reverberation mapping has been extended to the X-ray corona, accretion disk, and dust torus~\cite{Cackett2021}, it is most commonly applied to broad emission lines. The delay between variations in the continuum and broad-line emission, such as H$\beta$ $\lambda4861$,\AA, measures the characteristic size of the broad-line region. It is observationally challenging, requiring high-cadence monitoring over long baselines with minimal gaps. AGN variability is stochastic, so long campaigns are needed ($>3$ times the longest expected delay \cite{Horne2004}), and high signal-to-noise spectra are required to characterize variability at the few-percent level.

Most campaigns combine an emission-line delay with its spectral width and a population-averaged scale factor, accounting for the typically unknown geometry, kinematics, and inclination of the broad-line region~\cite{Onken2004,Batiste2017}, to constrain the enclosed mass. A growing number of campaigns include multiple emission lines, thus probing gas at different radii within the broad-line region (e.g. \cite{DeRosa2015,Williams2020,Homayouni2023,Bentz2021,Bentz2021a}). Data of exceptional quality allow for the full reverberation response as a function of line-of-sight velocity across a broad emission line to be constrained~\cite{Bentz2010}. Dynamical modeling can then constrain both the enclosed mass and the geometry and kinematics of the broad-line region, although currently the interpretation of such studies remains limited by modeling simplifications such as the lack of detailed photoionization physics~ \cite{Rosborough2025}.

\subsubsection{Mass measurements of dark matter spikes}
\label{sec:measurements_spikes}

While the study of DM spikes around super-massive BHs in AGN has remained limited, it has been pointed out that they may be tested observationally with reverberation mapping measurements~\cite{Sharma:2025ynw}. The idea is to exploit multi-line reverberation campaigns (Sec.~\ref{sec:measurements_MBH}): since different broad emission lines probe gas at different distances from the BH, they measure the mass enclosed within different radii, of order $10^{-3}$--$10^{-1}$~pc. A BH alone yields the same enclosed mass at all these radii, whereas a DM spike makes it grow with radius, see Fig.~\ref{fig:Spikes} (right). Applying this method to a sample of fourteen AGN, Ref.~\cite{Sharma:2025ynw} found that in five of them the enclosed mass increases with radius at the 1--2$\sigma$ level, with a preferred slope $\gamma_\mathrm{sp}\sim 1.6$, compatible with a spike mildly softened by stellar heating, although with a larger normalization than expected. These inferences are still limited by the large systematic uncertainties of current reverberation-mapping mass measurements (e.g.~in the virial factor and in the geometry of the broad-line region), strongly motivating further exploration of the connection between DM and AGN.
In addition to reverberation mapping, water megamaser disks have also been proposed to search for a population of dark matter spikes in distant AGN~\cite{Herrera:2026mao}.

DM spikes have received significantly larger attention for their impact on the gravitational waves of various BH inspirals~\cite{Bertone:2019irm}, see, e.g.,~\cite{Cole:2022yzw,Karydas:2024fcn,Kavanagh:2024lgq,DiCintio:2025ilt,Vicente:2025gsg} for some recent works.
Similar dynamical probes have been applied to supermassive BH binaries: the flare timings of the blazar OJ~287, which agree with general relativity up to 4.5 post-Newtonian order, constrain the mass of a spike around the primary BH to $\lesssim 3\%$ of its mass~\cite{Alachkar:2022oj287}. The observed orbital decay of the same system has been interpreted as evidence for a spike in agreement with the adiabatic expectation~\cite{Chan:2024oj287}, a claim which has been disputed in \cite{Deb:2025raq}.

Observations of Sgr A* provide an opportunity to probe DM spikes. Precision astrometric and spectroscopic measurements of stars orbiting Sgr A* constrain any extended mass distribution surrounding the black hole, thereby limiting the density of a possible DM spike \cite{GRAVITY:2024tth}. In addition, a DM spike would induce dynamical friction on celestial objects orbiting Sgr A* or other supermassive BHs, leading to potentially observable perturbations of their orbital motion \cite{Antonini:2011tu,Dosopoulou:2023umg}. The magnitude of these effects depends sensitively on the local DM density and hence are highly sensitive to feedback effects mentioned above. This is even more crucial in case further BSM interactions with DM particles are relevant, inducing an additional form of orbital energy loss in the celestial objects~\cite{Gustafson:2025ypo, Acevedo:2025rqu}.

\section{Path Forward}
\label{sec:questions}
AGN open an unprecedented window to the high-intensity and high-energy frontiers due to the large densities and energies of SM and DM particles around them,
none of which are attainable at Earth-based laboratory searches.
An important property of AGN is the availability of multi-messenger and multi-wavelength observations, which may be critical to disentangle astrophysical backgrounds from BSM-induced signals.

While AGN have a striking potential to deliver novel lessons about BSM (Sec.~\ref{sec:AGN_BSM}), the realisation of this potential requires improvements in the knowledge of these systems as well as of the DM around them. Both directions have recently seen critical progress, for reasons independent of the topic of this white paper. This makes it timely not only to review this progress through the lens of the AGN-BSM interplay (Secs~\ref{sec:AGN} and~\ref{sec:spikes}), but also to lay out the future steps that, in our opinion, are needed to derive solid BSM lessons from AGN.
In order to place robust bounds on BSM physics and potentially open the path to its discovery, the following three set of questions seem central to us:

\begin{enumerate}

\item[$\diamond$]
What can we learn about the particle acceleration mechanisms in AGN, and in particular about the luminosities and energies involved and about the morphology of the acceleration sites? How to maximise that learning from both the observational and the modelling points of view? 

\item[$\diamond$] What can we learn about the distribution of DM in the vicinity of supermassive BHs in AGN, both from observations and from simulations?
To what extent can we address these questions for individual sources? And statistically for an ensemble of sources? 

\item[$\diamond$] Which AGN observations could result in a BSM signal that is convincing, despite the uncertainties and unknowns about both AGN and DM around them? What BSM limits can be cast in a robust way from AGN observations?

\end{enumerate}
 Addressing these questions requires an interdisciplinary effort, bridging communities with expertise ranging from observational astronomy to theoretical astrophysics and particle physics. The purpose of this white paper is to launch this new perspective, and pave the way towards learning solid lessons about BSM physics from AGN.

\subsection*{Author contributions and AI disclosure}

G.H., F.S.\ and P.U.\ conceived and organized the workshop that led to this white paper, coordinated its preparation, and contributed equally to this document. All authors contributed crucially to discussions during and after the workshop that shaped the structure and content of this document, and wrote substantial parts of the text. Generative AI tools were used to assist with editing of limited parts of the text and with checking bibliographic references.

\subsection*{Acknowledgments}

The workshop organizers are thankful to IFPU for financial support of the
focus week program “Active Galactic Nuclei as Dark Sector Laboratories”
which inspired this publication. 
The work of G.H. was supported by the Neutrino Theory Network Fellowship with contract number 726844.
FS acknowledges funding by the European Union under the Horizon Europe’s Project: 101201278 – DarkSHunt - ERC - 2024 AdG and, together with AGDM, by the Italian INFN program on Theoretical Astroparticle Physics (TAsP). 
MCB gratefully acknowledges support from the NSF through grant AST-2407802. 
MC acknowledges support from the Spanish Agencia Estatal de Investigaci\'on through the grant PID2024-
155874NB-C22.
AG acknowledges funding from the Spanish grant PID2023-148162NB-C21, MCIN/AEI/10.13039/501100011033, co-funded by the European Union, FEDER. 
RAG is supported by DOE award DE-SC0020262 and by TRIUMF. TRIUMF receives federal funding via a
contribution agreement with the National Research Council (NRC) of Canada.
BJK acknowledges funding from the \textit{Consolidaci\'on Investigadora} Project \textsc{DarkSpikesGW}, reference CNS2023-144071, financed by MCIN/AEI/10.13039/501100011033 and by the European Union ``NextGenerationEU”/PRTR; and the project \textsc{CosmoRadio} (PID2025-169666NB-I00) financed by MICIU/AEI/10.13039/501100011033 and by the FSE+. The Instituto de Física de Cantabria (IFCA) has been accredited as Unidad de Excelencia María de Maeztu, a distinction awarded by the Spanish State Research Agency for the period 01/06/2026 - 31/05/2030.

\footnotesize
\bibliographystyle{JHEP}
\bibliography{References}

\providecommand{\href}[2]{#2}\begingroup\raggedright\begin{thebibliography}{100}

\bibitem{Cirelli:2024ssz}
M.~Cirelli, A.~Strumia and J.~Zupan, \emph{{Dark Matter}}, \href{http://dx.doi.org/10.21468/SciPostPhysRev.1}{\emph{SciPost Phys. Rev.} {\bf 1} (2026) }, [\href{http://arxiv.org/abs/2406.01705}{{\tt 2406.01705}}].

\bibitem{Padovani:2017zpf}
P.~Padovani et~al., \emph{{Active galactic nuclei: what{\textquoteright}s in a name?}}, \href{http://dx.doi.org/10.1007/s00159-017-0102-9}{\emph{Astron. Astrophys. Rev.} {\bf 25} (2017) 2}, [\href{http://arxiv.org/abs/1707.07134}{{\tt 1707.07134}}].

\bibitem{Abramowicz:2011xu}
M.~A. Abramowicz and P.~C. Fragile, \emph{{Foundations of Black Hole Accretion Disk Theory}}, \href{http://dx.doi.org/10.12942/lrr-2013-1}{\emph{Living Rev. Rel.} {\bf 16} (2013) 1}, [\href{http://arxiv.org/abs/1104.5499}{{\tt 1104.5499}}].

\bibitem{IceCube:2018dnn}
{\scshape IceCube, Fermi-LAT, MAGIC, AGILE, ASAS-SN, HAWC, H.E.S.S., INTEGRAL, Kanata, Kiso, Kapteyn, Liverpool Telescope, Subaru, Swift NuSTAR, VERITAS, VLA/17B-403} collaboration, M.~G. Aartsen et~al., \emph{{Multimessenger observations of a flaring blazar coincident with high-energy neutrino IceCube-170922A}}, \href{http://dx.doi.org/10.1126/science.aat1378}{\emph{Science} {\bf 361} (2018) eaat1378}, [\href{http://arxiv.org/abs/1807.08816}{{\tt 1807.08816}}].

\bibitem{IceCube:2022der}
{\scshape IceCube} collaboration, R.~Abbasi et~al., \emph{{Evidence for neutrino emission from the nearby active galaxy NGC 1068}}, \href{http://dx.doi.org/10.1126/science.abg3395}{\emph{Science} {\bf 378} (2022) 538--543}, [\href{http://arxiv.org/abs/2211.09972}{{\tt 2211.09972}}].

\bibitem{AGN_Dark_Sector_Laboratories_2026}
``\href{https://harvard-neutrinos.rc.fas.harvard.edu/event/33/}{Active Galactic Nuclei as Dark Sector Laboratories}.'' 15--19 June 2026, IFPU focus week, Trieste, Italy.

\bibitem{Gondolo:1999ef}
P.~Gondolo and J.~Silk, \emph{{Dark matter annihilation at the galactic center}}, \href{http://dx.doi.org/10.1103/PhysRevLett.83.1719}{\emph{Phys. Rev. Lett.} {\bf 83} (1999) 1719--1722}, [\href{http://arxiv.org/abs/astro-ph/9906391}{{\tt astro-ph/9906391}}].

\bibitem{Bloom:1997vm}
E.~D. Bloom and J.~D. Wells, \emph{{Multi - GeV photons from electron - dark matter scattering near active galactic nuclei}}, \href{http://dx.doi.org/10.1103/PhysRevD.57.1299}{\emph{Phys. Rev. D} {\bf 57} (1998) 1299--1302}, [\href{http://arxiv.org/abs/astro-ph/9706085}{{\tt astro-ph/9706085}}].

\bibitem{Gorchtein:2010xa}
M.~Gorchtein, S.~Profumo and L.~Ubaldi, \emph{{Probing Dark Matter with AGN Jets}}, \href{http://dx.doi.org/10.1103/PhysRevD.82.083514}{\emph{Phys. Rev. D} {\bf 82} (2010) 083514}, [\href{http://arxiv.org/abs/1008.2230}{{\tt 1008.2230}}].

\bibitem{Huang:2011dg}
J.-R. Huang, A.~Rajaraman and T.~M.~P. Tait, \emph{{Collisions of Jets of Particles from Active Galactic Nuclei with Neutralino Dark Matter}}, \href{http://dx.doi.org/10.1088/1475-7516/2012/05/027}{\emph{JCAP} {\bf 05} (2012) 027}, [\href{http://arxiv.org/abs/1109.2587}{{\tt 1109.2587}}].

\bibitem{Chang:2012sk}
S.~Chang, Y.~Gao and M.~Spannowsky, \emph{{Enhanced Gamma Ray Signals in Cosmic Proton-Wimp Collisions Due to Hadronization}}, \href{http://dx.doi.org/10.1088/1475-7516/2012/11/053}{\emph{JCAP} {\bf 11} (2012) 053}, [\href{http://arxiv.org/abs/1210.1870}{{\tt 1210.1870}}].

\bibitem{Profumo:2013jeb}
S.~Profumo, L.~Ubaldi and M.~Gorchtein, \emph{{Gamma Rays from Cosmic-Ray Proton Scattering in AGN Jets: the Intra-Cluster Gas vastly outshines Dark Matter}}, \href{http://dx.doi.org/10.1088/1475-7516/2013/04/012}{\emph{JCAP} {\bf 04} (2013) 012}, [\href{http://arxiv.org/abs/1302.1915}{{\tt 1302.1915}}].

\bibitem{Cermeno:2022rni}
M.~Cerme{\~n}o, C.~Degrande and L.~Mantani, \emph{{Signatures of leptophilic t-channel dark matter from active galactic nuclei}}, \href{http://dx.doi.org/10.1103/PhysRevD.105.083019}{\emph{Phys. Rev. D} {\bf 105} (2022) 083019}, [\href{http://arxiv.org/abs/2201.07247}{{\tt 2201.07247}}].

\bibitem{Rodrigues:2023vbv}
X.~Rodrigues, V.~S. Paliya, S.~Garrappa, A.~Omeliukh, A.~Franckowiak and W.~Winter, \emph{{Leptohadronic multi-messenger modeling of 324 gamma-ray blazars}}, \href{http://dx.doi.org/10.1051/0004-6361/202347540}{\emph{Astron. Astrophys.} {\bf 681} (2024) A119}, [\href{http://arxiv.org/abs/2307.13024}{{\tt 2307.13024}}].

\bibitem{DeMarchi:2025uoo}
A.~G. De~Marchi, A.~Granelli, J.~Nava and F.~Sala, \emph{{Boosted dark matter versus dark matter-induced neutrinos from single and stacked blazars}}, \href{http://dx.doi.org/10.1007/JHEP12(2025)136}{\emph{JHEP} {\bf 12} (2025) 136}, [\href{http://arxiv.org/abs/2507.12278}{{\tt 2507.12278}}].

\bibitem{Wang:2021jic}
J.-W. Wang, A.~Granelli and P.~Ullio, \emph{{Direct Detection Constraints on Blazar-Boosted Dark Matter}}, \href{http://dx.doi.org/10.1103/PhysRevLett.128.221104}{\emph{Phys. Rev. Lett.} {\bf 128} (2022) 221104}, [\href{http://arxiv.org/abs/2111.13644}{{\tt 2111.13644}}].

\bibitem{Granelli:2022ysi}
A.~Granelli, P.~Ullio and J.-W. Wang, \emph{{Blazar-boosted dark matter at Super-Kamiokande}}, \href{http://dx.doi.org/10.1088/1475-7516/2022/07/013}{\emph{JCAP} {\bf 07} (2022) 013}, [\href{http://arxiv.org/abs/2202.07598}{{\tt 2202.07598}}].

\bibitem{Bringmann:2018cvk}
T.~Bringmann and M.~Pospelov, \emph{{Novel direct detection constraints on light dark matter}}, \href{http://dx.doi.org/10.1103/PhysRevLett.122.171801}{\emph{Phys. Rev. Lett.} {\bf 122} (2019) 171801}, [\href{http://arxiv.org/abs/1810.10543}{{\tt 1810.10543}}].

\bibitem{Ema:2018bih}
Y.~Ema, F.~Sala and R.~Sato, \emph{{Light Dark Matter at Neutrino Experiments}}, \href{http://dx.doi.org/10.1103/PhysRevLett.122.181802}{\emph{Phys. Rev. Lett.} {\bf 122} (2019) 181802}, [\href{http://arxiv.org/abs/1811.00520}{{\tt 1811.00520}}].

\bibitem{Essig:2022dfa}
R.~Essig et~al., \emph{{Snowmass2021 Cosmic Frontier: The landscape of low-threshold dark matter direct detection in the next decade}},  in \emph{{Snowmass 2021}}, 3, 2022.
\newblock \href{http://arxiv.org/abs/2203.08297}{{\tt 2203.08297}}.

\bibitem{Bhowmick:2022zkj}
S.~Bhowmick, D.~Ghosh and D.~Sachdeva, \emph{{Blazar boosted dark matter {\textemdash} direct detection constraints on {\ensuremath{\sigma}}e{\ensuremath{\chi}}: role of energy dependent cross sections}}, \href{http://dx.doi.org/10.1088/1475-7516/2023/07/039}{\emph{JCAP} {\bf 07} (2023) 039}, [\href{http://arxiv.org/abs/2301.00209}{{\tt 2301.00209}}].

\bibitem{Jeesun:2025gzt}
S.~Jeesun, \emph{{Blazar boosted ALP and vector portal dark matter confronting light mediator searches}}, \href{http://dx.doi.org/10.1103/PhysRevD.111.103022}{\emph{Phys. Rev. D} {\bf 111} (2025) 103022}, [\href{http://arxiv.org/abs/2501.11569}{{\tt 2501.11569}}].

\bibitem{Wang:2025ztb}
J.-W. Wang, \emph{{Blazar-boosted dark matter: Novel signatures via elastic and inelastic scattering}}, \href{http://dx.doi.org/10.1103/wr32-3g38}{\emph{Phys. Rev. D} {\bf 112} (2025) 055004}, [\href{http://arxiv.org/abs/2503.22105}{{\tt 2503.22105}}].

\bibitem{Gustafson:2025dff}
R.~A. Gustafson, G.~Herrera, M.~Mukhopadhyay, K.~Murase and I.~M. Shoemaker, \emph{{Cosmic-ray boosted inelastic dark matter from neutrino-emitting active galactic nuclei}}, \href{http://dx.doi.org/10.1088/1475-7516/2026/05/068}{\emph{JCAP} {\bf 05} (2026) 068}, [\href{http://arxiv.org/abs/2508.20984}{{\tt 2508.20984}}].

\bibitem{Gago:2026ocx}
A.~M. Gago, J.~Hoefken~Zink, J.~Jones-P{\'e}rez and G.~D. Zapata, \emph{{Blazar Boosted Dark Matter in IceCube}},  \href{http://arxiv.org/abs/2608.14549}{{\tt 2608.14549}}.

\bibitem{Xia:2024ryt}
C.~Xia, C.-Y. Xing and Y.-H. Xu, \emph{{Boosted dark matter from Centaurus A and its detection}}, \href{http://dx.doi.org/10.1007/JHEP03(2024)076}{\emph{JHEP} {\bf 03} (2024) 076}, [\href{http://arxiv.org/abs/2401.03772}{{\tt 2401.03772}}].

\bibitem{Barillier:2025xct}
E.~Barillier, L.~Manenti, K.~Mora, P.~Padovani, I.~Sarnoff, Y.~Xu et~al., \emph{{Setting limits on blazar-boosted dark matter with xenon-based detectors}}, \href{http://dx.doi.org/10.1103/qxkb-4bpy}{\emph{Phys. Rev. D} {\bf 113} (2026) 023005}, [\href{http://arxiv.org/abs/2509.07265}{{\tt 2509.07265}}].

\bibitem{Cline:2022qld}
J.~M. Cline, S.~Gao, F.~Guo, Z.~Lin, S.~Liu, M.~Puel et~al., \emph{{Blazar Constraints on Neutrino-Dark Matter Scattering}}, \href{http://dx.doi.org/10.1103/PhysRevLett.130.091402}{\emph{Phys. Rev. Lett.} {\bf 130} (2023) 091402}, [\href{http://arxiv.org/abs/2209.02713}{{\tt 2209.02713}}].

\bibitem{Ferrer:2022kei}
F.~Ferrer, G.~Herrera and A.~Ibarra, \emph{{New constraints on the dark matter-neutrino and dark matter-photon scattering cross sections from TXS 0506+056}}, \href{http://dx.doi.org/10.1088/1475-7516/2023/05/057}{\emph{JCAP} {\bf 05} (2023) 057}, [\href{http://arxiv.org/abs/2209.06339}{{\tt 2209.06339}}].

\bibitem{Cline:2023tkp}
J.~M. Cline and M.~Puel, \emph{{NGC 1068 constraints on neutrino-dark matter scattering}}, \href{http://dx.doi.org/10.1088/1475-7516/2023/06/004}{\emph{JCAP} {\bf 06} (2023) 004}, [\href{http://arxiv.org/abs/2301.08756}{{\tt 2301.08756}}].

\bibitem{Fujiwara:2023lsv}
M.~Fujiwara and G.~Herrera, \emph{{Tidal disruption events and dark matter scatterings with neutrinos and photons}}, \href{http://dx.doi.org/10.1016/j.physletb.2024.138573}{\emph{Phys. Lett. B} {\bf 851} (2024) 138573}, [\href{http://arxiv.org/abs/2312.11670}{{\tt 2312.11670}}].

\bibitem{Fujiwara:2024qos}
M.~Fujiwara, G.~Herrera and S.~Horiuchi, \emph{{Neutrino Diffusion within Dark Matter Spikes}},  \href{http://arxiv.org/abs/2412.00805}{{\tt 2412.00805}}.

\bibitem{Cline:2026stf}
J.~M. Cline and M.~Puel, \emph{{Reply to ''Revisiting bounds on neutrino dark matter interaction at spikes''}},  \href{http://arxiv.org/abs/2608.17236}{{\tt 2608.17236}}.

\bibitem{Tseng:2024akh}
P.-Y. Tseng and Y.-M. Yeh, \emph{{Phenomenology of neutrino-dark matter interaction in DSNB and AGN}}, \href{http://dx.doi.org/10.1088/1475-7516/2025/08/038}{\emph{JCAP} {\bf 08} (2025) 038}, [\href{http://arxiv.org/abs/2412.08537}{{\tt 2412.08537}}].

\bibitem{Farzan:2026wek}
Y.~Farzan, \emph{{Revisiting bounds on neutrino dark matter interaction at spikes}},  \href{http://arxiv.org/abs/2607.08904}{{\tt 2607.08904}}.

\bibitem{Chiang:2026dmc}
C.-W. Chiang and V.~Q. Tran, \emph{{Threshold effects of inelastic neutrino-dark-matter scattering in AGN spikes}},  \href{http://arxiv.org/abs/2610.35620}{{\tt 2610.35620}}.

\bibitem{Zapata:2025huq}
G.~D. Zapata, J.~Jones-P{\'e}rez and A.~M. Gago, \emph{{Bounds on neutrino-DM interactions from TXS~0506+056 neutrino outburst}}, \href{http://dx.doi.org/10.1088/1475-7516/2025/07/042}{\emph{JCAP} {\bf 07} (2025) 042}, [\href{http://arxiv.org/abs/2503.03823}{{\tt 2503.03823}}].

\bibitem{Herrera:2025gpm}
G.~Herrera, \emph{{Plausible indication of gamma-ray absorption by dark matter in NGC 1068}}, \href{http://dx.doi.org/10.1016/j.dark.2025.102156}{\emph{Phys. Dark Univ.} {\bf 50} (2025) 102156}, [\href{http://arxiv.org/abs/2504.21560}{{\tt 2504.21560}}].

\bibitem{Bertolez-Martinez:2025trs}
T.~Bert{\'o}lez-Mart{\'\i}nez, G.~Herrera, P.~Mart{\'\i}nez-Mirav{\'e} and J.~Terol~Calvo, \emph{{Highest-energy neutrino event constrains dark matter-neutrino interactions}}, \href{http://dx.doi.org/10.1103/ds8v-lvmw}{\emph{Phys. Rev. D} {\bf 113} (2026) 103052}, [\href{http://arxiv.org/abs/2506.08993}{{\tt 2506.08993}}].

\bibitem{Mondol:2025uuw}
R.~Mondol, S.~Bouri, A.~K. Saha and R.~Laha, \emph{{Road through Dark$\nu$ess: Probing dark matter-neutrino interactions using KM3-230213A}},  \href{http://arxiv.org/abs/2506.19910}{{\tt 2506.19910}}.

\bibitem{Esteban:2025wbv}
I.~Esteban and A.~Ibarra, \emph{{Attenuation of the ultra-high-energy neutrino flux by dark matter scatterings}}, \href{http://dx.doi.org/10.1088/1475-7516/2026/04/064}{\emph{JCAP} {\bf 04} (2026) 064}, [\href{http://arxiv.org/abs/2508.02869}{{\tt 2508.02869}}].

\bibitem{Dixit:2026zkv}
K.~Dixit, G.~Mohlabeng and S.~Razzaque, \emph{{Searching for dark matter signals with high energy astrophysical neutrinos in IceCube}}, \href{http://dx.doi.org/10.1103/9zgv-fg7g}{\emph{Phys. Rev. D} {\bf 114} (2026) 043048}, [\href{http://arxiv.org/abs/2602.06121}{{\tt 2602.06121}}].

\bibitem{Esteban:2026vzh}
I.~Esteban, A.~M. Gago, M.~C. Gonzalez-Garcia and G.~D. Zapata, \emph{{Exploring neutrino loss with difusse astrophysical neutrino fluxes}}, \href{http://dx.doi.org/10.1088/1475-7516/2026/10/020}{\emph{JCAP} {\bf 10} (2026) 020}, [\href{http://arxiv.org/abs/2605.13955}{{\tt 2605.13955}}].

\bibitem{Abbaslu:2026row}
S.~Abbaslu, Y.~Farzan and A.~Maleki, \emph{{Flavorful Interactions of AGN Neutrinos with Dark Matter Spike}},  \href{http://arxiv.org/abs/2607.08833}{{\tt 2607.08833}}.

\bibitem{Tseng:2026evn}
P.-Y. Tseng and Y.-M. Yeh, \emph{{AGN and DSNB Neutrino Oscillation in Dark Matter Background}},  \href{http://arxiv.org/abs/2609.21768}{{\tt 2609.21768}}.

\bibitem{Dev:2026kqu}
P.~S.~B. Dev, E.~Gaido, A.~Ibarra and Y.~Porto, \emph{{Resonant neutrino flavor conversion within dark matter spikes}},  \href{http://arxiv.org/abs/2609.26773}{{\tt 2609.26773}}.

\bibitem{Herrera:2023nww}
G.~Herrera and K.~Murase, \emph{{Probing light dark matter through cosmic-ray cooling in active galactic nuclei}}, \href{http://dx.doi.org/10.1103/PhysRevD.110.L011701}{\emph{Phys. Rev. D} {\bf 110} (2024) L011701}, [\href{http://arxiv.org/abs/2307.09460}{{\tt 2307.09460}}].

\bibitem{Gustafson:2024aom}
R.~A. Gustafson, G.~Herrera, M.~Mukhopadhyay, K.~Murase and I.~M. Shoemaker, \emph{{Cosmic-ray cooling in active galactic nuclei as a new probe of inelastic dark matter}}, \href{http://dx.doi.org/10.1103/5m57-vgt2}{\emph{Phys. Rev. D} {\bf 111} (2025) L121303}, [\href{http://arxiv.org/abs/2408.08947}{{\tt 2408.08947}}].

\bibitem{Mishra:2025juk}
A.~K. Mishra, N.~Liu and C.-T. Lu, \emph{{Probing Gauged $U(1)$ Sub-GeV Dark Matter via Cosmic Ray Cooling in Active Galactic Nuclei}}, \href{http://dx.doi.org/10.1016/j.dark.2025.102050}{\emph{Phys. Dark Univ.} {\bf 49} (2025) 102050}, [\href{http://arxiv.org/abs/2504.03409}{{\tt 2504.03409}}].

\bibitem{Li:2025zwg}
L.~Li, C.-T. Lu, A.~K. Mishra, L.~Su and L.~Wu, \emph{{Enhanced cosmic-ray cooling in AGN from dark matter deep inelastic scattering}}, \href{http://dx.doi.org/10.1103/53ns-mxv3}{\emph{Phys. Rev. D} {\bf 114} (2026) 095026}, [\href{http://arxiv.org/abs/2509.11906}{{\tt 2509.11906}}].

\bibitem{Chun:2026muh}
E.~J. Chun, S.~Mandal and A.~Roy, \emph{{Probing Inelastic Dark Matter via Cosmic-Ray Upscattering in NGC 1068}},  \href{http://arxiv.org/abs/2606.03487}{{\tt 2606.03487}}.

\bibitem{Kantzas:2025huu}
D.~Kantzas, F.~Calore and M.~Chianese, \emph{{Cosmic-ray cooling by dark matter in astrophysical jets}}, \href{http://dx.doi.org/10.1103/fbfw-ptcw}{\emph{Phys. Rev. D} {\bf 113} (2026) 023048}, [\href{http://arxiv.org/abs/2509.18850}{{\tt 2509.18850}}].

\bibitem{Meighen-Berger:2025hrq}
S.~A. Meighen-Berger, P.~S.~B. Dev and M.~Hostert, \emph{{New multimessenger probe of dark matter-nucleon interactions from ultrahigh energy cosmic ray acceleration}}, \href{http://dx.doi.org/10.1103/747j-mhcq}{\emph{Phys. Rev. D} {\bf 113} (2026) 103050}, [\href{http://arxiv.org/abs/2512.18093}{{\tt 2512.18093}}].

\bibitem{Hussein:2025llu}
A.~Hussein and G.~Herrera, \emph{{Dark Matter-Electron Interactions Alter the Luminosity and Spectral Index of M87}},  \href{http://arxiv.org/abs/2510.12877}{{\tt 2510.12877}}.

\bibitem{DeMarchi:2024riu}
A.~G. De~Marchi, A.~Granelli, J.~Nava and F.~Sala, \emph{{Did IceCube discover dark matter around blazars?}}, \href{http://dx.doi.org/10.1103/lwqt-p598}{\emph{Phys. Rev. D} {\bf 112} (2025) 043042}, [\href{http://arxiv.org/abs/2412.07861}{{\tt 2412.07861}}].

\bibitem{DeMarchi:2025xag}
A.~G. De~Marchi, A.~Granelli, J.~Nava and F.~Sala, \emph{{Diffuse astrophysical neutrinos from dark matter around blazars}}, \href{http://dx.doi.org/10.1016/j.physletb.2025.140015}{\emph{Phys. Lett. B} {\bf 871} (2025) 140015}, [\href{http://arxiv.org/abs/2506.06416}{{\tt 2506.06416}}].

\bibitem{Murase:2018iyl}
K.~Murase, F.~Oikonomou and M.~Petropoulou, \emph{{Blazar Flares as an Origin of High-Energy Cosmic Neutrinos?}}, \href{http://dx.doi.org/10.3847/1538-4357/aada00}{\emph{Astrophys. J.} {\bf 865} (2018) 124}, [\href{http://arxiv.org/abs/1807.04748}{{\tt 1807.04748}}].

\bibitem{Gao:2018mnu}
S.~Gao, A.~Fedynitch, W.~Winter and M.~Pohl, \emph{{Modelling the coincident observation of a high-energy neutrino and a bright blazar flare}}, \href{http://dx.doi.org/10.1038/s41550-018-0610-1}{\emph{Nature Astron.} {\bf 3} (2019) 88--92}, [\href{http://arxiv.org/abs/1807.04275}{{\tt 1807.04275}}].

\bibitem{Cerruti:2018tmc}
M.~Cerruti, A.~Zech, C.~Boisson, G.~Emery, S.~Inoue and J.~P. Lenain, \emph{{Leptohadronic single-zone models for the electromagnetic and neutrino emission of TXS 0506+056}}, \href{http://dx.doi.org/10.1093/mnrasl/sly210}{\emph{Mon. Not. Roy. Astron. Soc.} {\bf 483} (2019) L12--L16}, [\href{http://arxiv.org/abs/1807.04335}{{\tt 1807.04335}}].

\bibitem{Keivani:2018rnh}
A.~Keivani et~al., \emph{{A Multimessenger Picture of the Flaring Blazar TXS 0506+056: implications for High-Energy Neutrino Emission and Cosmic Ray Acceleration}}, \href{http://dx.doi.org/10.3847/1538-4357/aad59a}{\emph{Astrophys. J.} {\bf 864} (2018) 84}, [\href{http://arxiv.org/abs/1807.04537}{{\tt 1807.04537}}].

\bibitem{Akita:2025dhg}
K.~Akita, A.~Ibarra and R.~Zimmermann, \emph{{Dark matter explanations for the neutrino emission from the Seyfert galaxy NGC 1068}},  \href{http://arxiv.org/abs/2507.16539}{{\tt 2507.16539}}.

\bibitem{Ayad:2019hrj}
A.~Ayad and G.~Beck, \emph{{Probing a cosmic axion-like particle background within the jets of active galactic nuclei}}, \href{http://dx.doi.org/10.1088/1475-7516/2020/04/055}{\emph{JCAP} {\bf 04} (2020) 055}, [\href{http://arxiv.org/abs/1911.10078}{{\tt 1911.10078}}].

\bibitem{Davies:2022wvj}
J.~Davies, M.~Meyer and G.~Cotter, \emph{{Constraints on axionlike particles from a combined analysis of three flaring Fermi flat-spectrum radio quasars}}, \href{http://dx.doi.org/10.1103/PhysRevD.107.083027}{\emph{Phys. Rev. D} {\bf 107} (2023) 083027}, [\href{http://arxiv.org/abs/2211.03414}{{\tt 2211.03414}}].

\bibitem{DeMarchi:2026pak}
A.~G. De~Marchi, O.~Ning and T.~Xiao, \emph{{Blazar Constraints on Axions through New Spectral Modulation Searches in 1ES 1959+650 {\&} B2 1811+31}},  \href{http://arxiv.org/abs/2603.13480}{{\tt 2603.13480}}.

\bibitem{Chen:2017mjf}
L.~Chen and J.~P. Conlon, \emph{{Constraints on massive axion-like particles from X-ray observations of NGC 1275}}, \href{http://dx.doi.org/10.1093/mnras/sty1591}{\emph{Mon. Not. Roy. Astron. Soc.} {\bf 479} (2018) 2243--2248}, [\href{http://arxiv.org/abs/1712.08313}{{\tt 1712.08313}}].

\bibitem{Malyshev:2025iis}
D.~Malyshev, L.~Zadorozhna, Y.~Bidasyuk, A.~Santangelo and O.~Ruchayskiy, \emph{{Constraints on axion-like particles from active galactic nuclei seen through galaxy clusters}}, \href{http://dx.doi.org/10.1038/s41550-025-02621-8}{\emph{Nature Astron.} {\bf 9} (2025) 1387--1395}, [\href{http://arxiv.org/abs/2506.02848}{{\tt 2506.02848}}].

\bibitem{Grimaud:2026hpu}
C.~Grimaud, D.~Malyshev and E.~Moulin, \emph{{Sensitivity to axion-like particle dark matter with very-high-energy gamma-ray observations of active galactic nuclei located behind galaxy clusters}}, \href{http://dx.doi.org/10.1088/1475-7516/2026/07/066}{\emph{JCAP} {\bf 07} (2026) 066}, [\href{http://arxiv.org/abs/2603.12009}{{\tt 2603.12009}}].

\bibitem{Li:2020pcn}
H.-J. Li, J.-G. Guo, X.-J. Bi, S.-J. Lin and P.-F. Yin, \emph{{Limits on axion-like particles from Mrk 421 with 4.5-year period observations by ARGO-YBJ and Fermi-LAT}}, \href{http://dx.doi.org/10.1103/PhysRevD.103.083003}{\emph{Phys. Rev. D} {\bf 103} (2021) 083003}, [\href{http://arxiv.org/abs/2008.09464}{{\tt 2008.09464}}].

\bibitem{Yao:2022col}
R.-M. Yao, X.-J. Bi, J.-W. Wang and P.-F. Yin, \emph{{Optical circular polarization induced by axionlike particles in blazars}}, \href{http://dx.doi.org/10.1103/PhysRevD.107.043031}{\emph{Phys. Rev. D} {\bf 107} (2023) 043031}, [\href{http://arxiv.org/abs/2209.14214}{{\tt 2209.14214}}].

\bibitem{Dekker:2025vcg}
A.~Dekker, G.~Herrera and D.~Kantzas, \emph{{Axionlike particle limits from multimessenger sources}}, \href{http://dx.doi.org/10.1103/8mkk-g1ck}{\emph{Phys. Rev. D} {\bf 113} (2026) 123039}, [\href{http://arxiv.org/abs/2506.14659}{{\tt 2506.14659}}].

\bibitem{Dev:2025czz}
P.~S.~B. Dev, B.~Dutta, A.~Karthikeyan, W.~Maitra, L.~E. Strigari and A.~Verma, \emph{{`Dark' Matter Effect as a Novel Solution to the KM3-230213A Puzzle}},  \href{http://arxiv.org/abs/2505.22754}{{\tt 2505.22754}}.

\bibitem{Ettengruber:2025kat}
M.~Ettengruber and G.~Herrera, \emph{{New Tests of Low-Scale Quantum Gravity with Cosmic-Ray Collisions}},  \href{http://arxiv.org/abs/2510.11879}{{\tt 2510.11879}}.

\bibitem{Arguelles:2026rjr}
C.~A. Arg{\"u}elles, P.~S.~B. Dev, B.~Dutta, G.~Herrera, N.~Kamp, J.~Kumar et~al., \emph{{Astrophysical Neutrino Sources as Colliders}},  \href{http://arxiv.org/abs/2607.19254}{{\tt 2607.19254}}.

\bibitem{Arvanitaki:2014wva}
A.~Arvanitaki, M.~Baryakhtar and X.~Huang, \emph{{Discovering the QCD Axion with Black Holes and Gravitational Waves}}, \href{http://dx.doi.org/10.1103/PhysRevD.91.084011}{\emph{Phys. Rev. D} {\bf 91} (2015) 084011}, [\href{http://arxiv.org/abs/1411.2263}{{\tt 1411.2263}}].

\bibitem{Unal:2020jiy}
C.~{\"U}nal, F.~Pacucci and A.~Loeb, \emph{{Properties of ultralight bosons from heavy quasar spins via superradiance}}, \href{http://dx.doi.org/10.1088/1475-7516/2021/05/007}{\emph{JCAP} {\bf 05} (2021) 007}, [\href{http://arxiv.org/abs/2012.12790}{{\tt 2012.12790}}].

\bibitem{Cardoso:2018tly}
V.~Cardoso, {\'O}.~J.~C. Dias, G.~S. Hartnett, M.~Middleton, P.~Pani and J.~E. Santos, \emph{{Constraining the mass of dark photons and axion-like particles through black-hole superradiance}}, \href{http://dx.doi.org/10.1088/1475-7516/2018/03/043}{\emph{JCAP} {\bf 03} (2018) 043}, [\href{http://arxiv.org/abs/1801.01420}{{\tt 1801.01420}}].

\bibitem{Baryakhtar:2017ngi}
M.~Baryakhtar, R.~Lasenby and M.~Teo, \emph{{Black Hole Superradiance Signatures of Ultralight Vectors}}, \href{http://dx.doi.org/10.1103/PhysRevD.96.035019}{\emph{Phys. Rev. D} {\bf 96} (2017) 035019}, [\href{http://arxiv.org/abs/1704.05081}{{\tt 1704.05081}}].

\bibitem{Brito:2017wnc}
R.~Brito, S.~Ghosh, E.~Barausse, E.~Berti, V.~Cardoso, I.~Dvorkin et~al., \emph{{Stochastic and resolvable gravitational waves from ultralight bosons}}, \href{http://dx.doi.org/10.1103/PhysRevLett.119.131101}{\emph{Phys. Rev. Lett.} {\bf 119} (2017) 131101}, [\href{http://arxiv.org/abs/1706.05097}{{\tt 1706.05097}}].

\bibitem{Chen:2019fsq}
Y.~Chen, J.~Shu, X.~Xue, Q.~Yuan and Y.~Zhao, \emph{{Probing Axions with Event Horizon Telescope Polarimetric Measurements}}, \href{http://dx.doi.org/10.1103/PhysRevLett.124.061102}{\emph{Phys. Rev. Lett.} {\bf 124} (2020) 061102}, [\href{http://arxiv.org/abs/1905.02213}{{\tt 1905.02213}}].

\bibitem{Siemonsen:2022ivj}
N.~Siemonsen, C.~Mondino, D.~Egana-Ugrinovic, J.~Huang, M.~Baryakhtar and W.~E. East, \emph{{Dark photon superradiance: Electrodynamics and multimessenger signals}}, \href{http://dx.doi.org/10.1103/PhysRevD.107.075025}{\emph{Phys. Rev. D} {\bf 107} (2023) 075025}, [\href{http://arxiv.org/abs/2212.09772}{{\tt 2212.09772}}].

\bibitem{Sarmah:2024nst}
P.~Sarmah, H.~Verma, K.~Cheung and J.~Silk, \emph{{Effects of superradiance in active galactic nuclei}}, \href{http://dx.doi.org/10.1093/mnras/staf326}{\emph{Mon. Not. Roy. Astron. Soc.} {\bf 538} (2025) 943--962}, [\href{http://arxiv.org/abs/2404.09955}{{\tt 2404.09955}}].

\bibitem{IceCube:2018cha}
{\scshape IceCube} collaboration, M.~G. Aartsen et~al., \emph{{Neutrino emission from the direction of the blazar TXS 0506+056 prior to the IceCube-170922A alert}}, \href{http://dx.doi.org/10.1126/science.aat2890}{\emph{Science} {\bf 361} (2018) 147--151}, [\href{http://arxiv.org/abs/1807.08794}{{\tt 1807.08794}}].

\bibitem{MAGIC:2022gyf}
{\scshape MAGIC, OVRO, Mets{\"a}hovi} collaboration, V.~A. Acciari et~al., \emph{{Investigating the Blazar TXS 0506+056 through Sharp Multiwavelength Eyes During 2017{\textendash}2019}}, \href{http://dx.doi.org/10.3847/1538-4357/ac531d}{\emph{Astrophys. J.} {\bf 927} (2022) 197}, [\href{http://arxiv.org/abs/2202.02600}{{\tt 2202.02600}}].

\bibitem{Fermi-LAT:2022byn}
{\scshape Fermi-LAT} collaboration, S.~Abdollahi et~al., \emph{{Incremental Fermi Large Area Telescope Fourth Source Catalog}}, \href{http://dx.doi.org/10.3847/1538-4365/ac6751}{\emph{Astrophys. J. Supp.} {\bf 260} (2022) 53}, [\href{http://arxiv.org/abs/2201.11184}{{\tt 2201.11184}}].

\bibitem{Garc_a_Burillo_2016}
S.~García-Burillo, F.~Combes, C.~R. Almeida, A.~Usero, M.~Krips, A.~Alonso-Herrero et~al., \emph{Alma resolves the torus of ngc 1068: Continuum and molecular line emission}, \href{http://dx.doi.org/10.3847/2041-8205/823/1/l12}{\emph{The Astrophysical Journal Letters} {\bf 823} (May, 2016) L12}.

\bibitem{Impellizzeri_2019}
C.~M.~V. Impellizzeri, J.~F. Gallimore, S.~A. Baum, M.~Elitzur, R.~Davies, D.~Lutz et~al., \emph{Counter-rotation and high-velocity outflow in the parsec-scale molecular torus of ngc 1068}, \href{http://dx.doi.org/10.3847/2041-8213/ab3c64}{\emph{The Astrophysical Journal Letters} {\bf 884} (Oct., 2019) L28}.

\bibitem{BERESINSKY1969423}
V.~Beresinsky and G.~Zatsepin, \emph{Cosmic rays at ultra high energies (neutrino?)}, \href{http://dx.doi.org/https://doi.org/10.1016/0370-2693(69)90341-4}{\emph{Physics Letters B} {\bf 28} (1969) 423--424}.

\bibitem{Berezinsky:1975zz}
V.~S. Berezinsky and A.~Y. Smirnov, \emph{{Cosmic neutrinos of ultra-high energies and detection possibility}}, \href{http://dx.doi.org/10.1007/BF00643157}{\emph{Astrophys. Space Sci.} {\bf 32} (1975) 461--482}.

\bibitem{Eichler:1979yy}
D.~Eichler, \emph{{High-energy neutrino astronomy: a probe of galactic nuclei?}}, \href{https://ui.adsabs.harvard.edu/abs/1979ApJ...232..106E}{\emph{Astrophys. J.} {\bf 232} (1979) 106--112}.

\bibitem{Berezinsky:1981sdl}
V.~S. Berezinsky and V.~L. Ginzburg, \emph{{On high-energy neutrino radiation of quasars and active galactic nuclei}}, \href{http://dx.doi.org/10.1093/mnras/194.1.3}{\emph{Mon. Not. Roy. Astron. Soc.} {\bf 194} (1981) 3--14}.

\bibitem{Stecker:1991vm}
F.~W. Stecker, C.~Done, M.~H. Salamon and P.~Sommers, \emph{{High-energy neutrinos from active galactic nuclei}}, \href{http://dx.doi.org/10.1103/PhysRevLett.66.2697}{\emph{Phys. Rev. Lett.} {\bf 66} (1991) 2697--2700}.

\bibitem{Murase:2022feu}
K.~Murase and F.~W. Stecker, \emph{{Chapter 10: High-Energy Neutrinos from Active Galactic Nuclei}},  \href{http://arxiv.org/abs/2202.03381}{{\tt 2202.03381}}.

\bibitem{Rodrigues:2025cpm}
X.~Rodrigues, F.~Rieger, A.~Bohdan and P.~Padovani, \emph{{Hillas meets Eddington: The case for blazars as ultra-high-energy neutrino sources}}, \href{http://dx.doi.org/10.1051/0004-6361/202556986}{\emph{Astron. Astrophys.} {\bf 706} (2026) A351}, [\href{http://arxiv.org/abs/2508.18345}{{\tt 2508.18345}}].

\bibitem{KM3NeT:2025npi}
{\scshape KM3NeT} collaboration, S.~Aiello et~al., \emph{{Observation of an ultra-high-energy cosmic neutrino with KM3NeT}}, \href{http://dx.doi.org/10.1038/s41586-024-08543-1}{\emph{Nature} {\bf 638} (2025) 376--382}.

\bibitem{Urry:1995mg}
C.~M. Urry and P.~Padovani, \emph{{Unified schemes for radio-loud active galactic nuclei}}, \href{http://dx.doi.org/10.1086/133630}{\emph{Publ. Astron. Soc. Pac.} {\bf 107} (1995) 803}, [\href{http://arxiv.org/abs/astro-ph/9506063}{{\tt astro-ph/9506063}}].

\bibitem{Ballet:2026kpq}
{\scshape Fermi-LAT} collaboration, J.~Ballet, P.~Bruel, T.~H. Burnett and B.~Lott, \emph{{Fermi-LAT 16-year Source List}},  \href{http://arxiv.org/abs/2602.22148}{{\tt 2602.22148}}.

\bibitem{CTAConsortium:2017dvg}
{\scshape CTA Consortium} collaboration, B.~S. Acharya et~al., \emph{{Science with the Cherenkov Telescope Array}}.
\newblock WSP, 11, 2018, \href{http://dx.doi.org/10.1142/10986}{10.1142/10986}.

\bibitem{DiMauro:2013xta}
M.~Di~Mauro, F.~Calore, F.~Donato, M.~Ajello and L.~Latronico, \emph{{Diffuse $\gamma$-ray emission from misaligned active galactic nuclei}}, \href{http://dx.doi.org/10.1088/0004-637X/780/2/161}{\emph{Astrophys. J.} {\bf 780} (2014) 161}, [\href{http://arxiv.org/abs/1304.0908}{{\tt 1304.0908}}].

\bibitem{Ajello:2011zi}
M.~Ajello et~al., \emph{{The Luminosity Function of Fermi-detected Flat-Spectrum Radio Quasars}}, \href{http://dx.doi.org/10.1088/0004-637X/751/2/108}{\emph{Astrophys. J.} {\bf 751} (2012) 108}, [\href{http://arxiv.org/abs/1110.3787}{{\tt 1110.3787}}].

\bibitem{Ajello:2013lka}
M.~Ajello et~al., \emph{{The Cosmic Evolution of Fermi BL Lacertae Objects}}, \href{http://dx.doi.org/10.1088/0004-637X/780/1/73}{\emph{Astrophys. J.} {\bf 780} (2014) 73}, [\href{http://arxiv.org/abs/1310.0006}{{\tt 1310.0006}}].

\bibitem{Qu:2019zln}
Y.~Qu, H.~Zeng and D.~Yan, \emph{{Gamma-ray luminosity function of BL Lac objects and contribution to the extragalactic gamma-ray background}}, \href{http://dx.doi.org/10.1093/mnras/stz2651}{\emph{Mon. Not. Roy. Astron. Soc.} {\bf 490} (2019) 758--765}, [\href{http://arxiv.org/abs/1909.07542}{{\tt 1909.07542}}].

\bibitem{Blanco:2021icw}
C.~Blanco and T.~Linden, \emph{{Star-forming galaxies provide a larger contribution to the isotropic gamma-ray background than misaligned active galactic nuclei}}, \href{http://dx.doi.org/10.1088/1475-7516/2023/02/003}{\emph{JCAP} {\bf 02} (2023) 003}, [\href{http://arxiv.org/abs/2104.03315}{{\tt 2104.03315}}].

\bibitem{Murase:2012df}
K.~Murase, J.~F. Beacom and H.~Takami, \emph{{Gamma-Ray and Neutrino Backgrounds as Probes of the High-Energy Universe: Hints of Cascades, General Constraints, and Implications for TeV Searches}}, \href{http://dx.doi.org/10.1088/1475-7516/2012/08/030}{\emph{JCAP} {\bf 08} (2012) 030}, [\href{http://arxiv.org/abs/1205.5755}{{\tt 1205.5755}}].

\bibitem{Yan:2012xs}
D.~Yan, H.~Zeng and L.~Zhang, \emph{{The contribution from blazar cascade emission to the extragalactic gamma-ray background: What a role does the extragalactic magnetic field play?}}, \href{http://dx.doi.org/10.1111/j.1365-2966.2012.20752.x}{\emph{Mon. Not. Roy. Astron. Soc.} {\bf 422} (2012) 1779--1784}, [\href{http://arxiv.org/abs/1202.4250}{{\tt 1202.4250}}].

\bibitem{Blanco:2023kfa}
C.~Blanco, O.~Ghosh, S.~Jacobsen and T.~Linden, \emph{{Where are the Cascades from Blazar Jets? An Emerging Tension in the $\gamma$-ray sky}},  \href{http://arxiv.org/abs/2303.01524}{{\tt 2303.01524}}.

\bibitem{Fermi-LAT:2015qzw}
{\scshape Fermi-LAT} collaboration, M.~Ackermann et~al., \emph{{Limits on Dark Matter Annihilation Signals from the Fermi LAT 4-year Measurement of the Isotropic Gamma-Ray Background}}, \href{http://dx.doi.org/10.1088/1475-7516/2015/09/008}{\emph{JCAP} {\bf 09} (2015) 008}, [\href{http://arxiv.org/abs/1501.05464}{{\tt 1501.05464}}].

\bibitem{Murase:2012xs}
K.~Murase and J.~F. Beacom, \emph{{Constraining Very Heavy Dark Matter Using Diffuse Backgrounds of Neutrinos and Cascaded Gamma Rays}}, \href{http://dx.doi.org/10.1088/1475-7516/2012/10/043}{\emph{JCAP} {\bf 10} (2012) 043}, [\href{http://arxiv.org/abs/1206.2595}{{\tt 1206.2595}}].

\bibitem{Murase:2015gea}
K.~Murase, R.~Laha, S.~Ando and M.~Ahlers, \emph{{Testing the Dark Matter Scenario for PeV Neutrinos Observed in IceCube}}, \href{http://dx.doi.org/10.1103/PhysRevLett.115.071301}{\emph{Phys. Rev. Lett.} {\bf 115} (2015) 071301}, [\href{http://arxiv.org/abs/1503.04663}{{\tt 1503.04663}}].

\bibitem{Blanco:2018esa}
C.~Blanco and D.~Hooper, \emph{{Constraints on Decaying Dark Matter from the Isotropic Gamma-Ray Background}}, \href{http://dx.doi.org/10.1088/1475-7516/2019/03/019}{\emph{JCAP} {\bf 03} (2019) 019}, [\href{http://arxiv.org/abs/1811.05988}{{\tt 1811.05988}}].

\bibitem{Das:2023wtk}
S.~Das, K.~Murase and T.~Fujii, \emph{{Revisiting ultrahigh-energy constraints on decaying superheavy dark matter}}, \href{http://dx.doi.org/10.1103/PhysRevD.107.103013}{\emph{Phys. Rev. D} {\bf 107} (2023) 103013}, [\href{http://arxiv.org/abs/2302.02993}{{\tt 2302.02993}}].

\bibitem{Aartsen_2017}
M.~Aartsen, M.~Ackermann, J.~Adams, J.~Aguilar, M.~Ahlers, M.~Ahrens et~al., \emph{The icecube neutrino observatory: instrumentation and online systems}, \href{http://dx.doi.org/10.1088/1748-0221/12/03/P03012}{\emph{Journal of Instrumentation} {\bf 12} (mar, 2017) P03012}.

\bibitem{Abbasi_2026}
R.~Abbasi, M.~Ackermann, J.~Adams, S.~K. Agarwalla, J.~A. Aguilar, M.~Ahlers et~al., \emph{Evidence for neutrino emission from x-ray-bright active galactic nuclei with icecube}, \href{http://dx.doi.org/10.3847/2041-8213/ae4aad}{\emph{The Astrophysical Journal Letters} {\bf 1000} (mar, 2026) L26}.

\bibitem{IceCube:2024ayt}
{\scshape IceCube} collaboration, R.~Abbasi et~al., \emph{{Search for Neutrino Emission from Hard X-Ray AGN with IceCube}}, \href{http://dx.doi.org/10.3847/1538-4357/ada94b}{\emph{Astrophys. J.} {\bf 981} (2025) 131}, [\href{http://arxiv.org/abs/2406.06684}{{\tt 2406.06684}}].

\bibitem{IceCube:2024dou}
{\scshape IceCube} collaboration, R.~Abbasi et~al., \emph{{IceCube Search for Neutrino Emission from X-Ray Bright Seyfert Galaxies}}, \href{http://dx.doi.org/10.3847/1538-4357/addd05}{\emph{Astrophys. J.} {\bf 988} (2025) 141}, [\href{http://arxiv.org/abs/2406.07601}{{\tt 2406.07601}}].

\bibitem{Abbasi_2026_2}
R.~Abbasi, M.~Ackermann, J.~Adams, S.~K. Agarwalla, J.~A. Aguilar, M.~Ahlers et~al., \emph{Evidence for neutrino emission from x-ray bright seyfert galaxies in the southern hemisphere using enhanced starting track events with icecube}, \href{http://dx.doi.org/10.3847/2041-8213/ae4aac}{\emph{The Astrophysical Journal Letters} {\bf 1000} (mar, 2026) L37}.

\bibitem{doi:10.1126/science.aat2890}
I.~Collaboration, M.~Aartsen, M.~Ackermann, J.~Adams, J.~A. Aguilar, M.~Ahlers et~al., \emph{Neutrino emission from the direction of the blazar txs 0506+056 prior to the icecube-170922a alert}, \href{http://dx.doi.org/10.1126/science.aat2890}{\emph{Science} {\bf 361} (2018) 147--151}, [\href{http://arxiv.org/abs/https://www.science.org/doi/pdf/10.1126/science.aat2890}{{\tt https://www.science.org/doi/pdf/10.1126/science.aat2890}}].

\bibitem{IceCube:2025tmc}
{\scshape IceCube} collaboration, S.~Hori et~al., \emph{{IceCube population constraints on neutrino emission by Fermi-LAT detected active galactic nuclei}}, \href{http://dx.doi.org/10.22323/1.501.0933}{\emph{PoS} {\bf ICRC2025} (2025) 933}, [\href{http://arxiv.org/abs/2507.07098}{{\tt 2507.07098}}].

\bibitem{Abbasi_2023}
R.~Abbasi, M.~Ackermann, J.~Adams, S.~K. Agarwalla, J.~A. Aguilar, M.~Ahlers et~al., \emph{Search for correlations of high-energy neutrinos detected in icecube with radio-bright agn and gamma-ray emission from blazars}, \href{http://dx.doi.org/10.3847/1538-4357/acdfcb}{\emph{The Astrophysical Journal} {\bf 954} (aug, 2023) 75}.

\bibitem{IceCube:2016qvd}
{\scshape IceCube} collaboration, M.~G. Aartsen et~al., \emph{{The contribution of Fermi-2LAC blazars to the diffuse TeV-PeV neutrino flux}}, \href{http://dx.doi.org/10.3847/1538-4357/835/1/45}{\emph{Astrophys. J.} {\bf 835} (2017) 45}, [\href{http://arxiv.org/abs/1611.03874}{{\tt 1611.03874}}].

\bibitem{Murase:2013rfa}
K.~Murase, M.~Ahlers and B.~C. Lacki, \emph{{Testing the Hadronuclear Origin of PeV Neutrinos Observed with IceCube}}, \href{http://dx.doi.org/10.1103/PhysRevD.88.121301}{\emph{Phys. Rev. D} {\bf 88} (2013) 121301}, [\href{http://arxiv.org/abs/1306.3417}{{\tt 1306.3417}}].

\bibitem{Murase:2015xka}
K.~Murase, D.~Guetta and M.~Ahlers, \emph{{Hidden Cosmic-Ray Accelerators as an Origin of TeV-PeV Cosmic Neutrinos}}, \href{http://dx.doi.org/10.1103/PhysRevLett.116.071101}{\emph{Phys. Rev. Lett.} {\bf 116} (2016) 071101}, [\href{http://arxiv.org/abs/1509.00805}{{\tt 1509.00805}}].

\bibitem{Bechtol:2015uqb}
K.~Bechtol, M.~Ahlers, M.~Di~Mauro, M.~Ajello and J.~Vandenbroucke, \emph{{Evidence against star-forming galaxies as the dominant source of IceCube neutrinos}}, \href{http://dx.doi.org/10.3847/1538-4357/836/1/47}{\emph{Astrophys. J.} {\bf 836} (2017) 47}, [\href{http://arxiv.org/abs/1511.00688}{{\tt 1511.00688}}].

\bibitem{Fermi-LAT:2015otn}
{\scshape Fermi-LAT} collaboration, M.~Ackermann et~al., \emph{{Resolving the Extragalactic $\gamma$-Ray Background above 50 GeV with the Fermi Large Area Telescope}}, \href{http://dx.doi.org/10.1103/PhysRevLett.116.151105}{\emph{Phys. Rev. Lett.} {\bf 116} (2016) 151105}, [\href{http://arxiv.org/abs/1511.00693}{{\tt 1511.00693}}].

\bibitem{Lisanti:2016jub}
M.~Lisanti, S.~Mishra-Sharma, L.~Necib and B.~R. Safdi, \emph{{Deciphering Contributions to the Extragalactic Gamma-Ray Background from 2 GeV to 2 TeV}}, \href{http://dx.doi.org/10.3847/0004-637X/832/2/117}{\emph{Astrophys. J.} {\bf 832} (2016) 117}, [\href{http://arxiv.org/abs/1606.04101}{{\tt 1606.04101}}].

\bibitem{Ajello:2015mfa}
M.~Ajello et~al., \emph{{The Origin of the Extragalactic Gamma-Ray Background and Implications for Dark-Matter Annihilation}}, \href{http://dx.doi.org/10.1088/2041-8205/800/2/L27}{\emph{Astrophys. J. Lett.} {\bf 800} (2015) L27}, [\href{http://arxiv.org/abs/1501.05301}{{\tt 1501.05301}}].

\bibitem{Hooper:2018wyk}
D.~Hooper, T.~Linden and A.~Vieregg, \emph{{Active Galactic Nuclei and the Origin of IceCube's Diffuse Neutrino Flux}}, \href{http://dx.doi.org/10.1088/1475-7516/2019/02/012}{\emph{JCAP} {\bf 02} (2019) 012}, [\href{http://arxiv.org/abs/1810.02823}{{\tt 1810.02823}}].

\bibitem{Capanema:2020rjj}
A.~Capanema, A.~Esmaili and K.~Murase, \emph{{New constraints on the origin of medium-energy neutrinos observed by IceCube}}, \href{http://dx.doi.org/10.1103/PhysRevD.101.103012}{\emph{Phys. Rev. D} {\bf 101} (2020) 103012}, [\href{http://arxiv.org/abs/2002.07192}{{\tt 2002.07192}}].

\bibitem{Inoue:2019yfs}
Y.~Inoue, D.~Khangulyan and A.~Doi, \emph{{On the Origin of High-energy Neutrinos from NGC 1068: The Role of Nonthermal Coronal Activity}}, \href{http://dx.doi.org/10.3847/2041-8213/ab7661}{\emph{Astrophys. J. Lett.} {\bf 891} (2020) L33}, [\href{http://arxiv.org/abs/1909.02239}{{\tt 1909.02239}}].

\bibitem{RNO-G:2020rmc}
{\scshape RNO-G} collaboration, J.~A. Aguilar et~al., \emph{{Design and Sensitivity of the Radio Neutrino Observatory in Greenland (RNO-G)}}, \href{http://dx.doi.org/10.1088/1748-0221/16/03/P03025}{\emph{JINST} {\bf 16} (2021) P03025}, [\href{http://arxiv.org/abs/2010.12279}{{\tt 2010.12279}}].

\bibitem{IceCube-Gen2:2020qha}
{\scshape IceCube-Gen2} collaboration, M.~G. Aartsen et~al., \emph{{IceCube-Gen2: the window to the extreme Universe}}, \href{http://dx.doi.org/10.1088/1361-6471/abbd48}{\emph{J. Phys. G} {\bf 48} (2021) 060501}, [\href{http://arxiv.org/abs/2008.04323}{{\tt 2008.04323}}].

\bibitem{Murase:2019vdl}
K.~Murase, S.~S. Kimura and P.~Meszaros, \emph{{Hidden Cores of Active Galactic Nuclei as the Origin of Medium-Energy Neutrinos: Critical Tests with the MeV Gamma-Ray Connection}}, \href{http://dx.doi.org/10.1103/PhysRevLett.125.011101}{\emph{Phys. Rev. Lett.} {\bf 125} (2020) 011101}, [\href{http://arxiv.org/abs/1904.04226}{{\tt 1904.04226}}].

\bibitem{Ajello:2023hkh}
M.~Ajello, K.~Murase and A.~McDaniel, \emph{{Disentangling the Hadronic Components in NGC 1068}}, \href{http://dx.doi.org/10.3847/2041-8213/acf296}{\emph{Astrophys. J. Lett.} {\bf 954} (2023) L49}, [\href{http://arxiv.org/abs/2307.02333}{{\tt 2307.02333}}].

\bibitem{Cerruti:2020lfj}
M.~Cerruti, \emph{{Leptonic and Hadronic Radiative Processes in Supermassive-Black-Hole Jets}}, \href{http://dx.doi.org/10.3390/galaxies8040072}{\emph{Galaxies} {\bf 8} (2020) 72}, [\href{http://arxiv.org/abs/2012.13302}{{\tt 2012.13302}}].

\bibitem{Rodrigues:2026kpf}
X.~Rodrigues, \emph{{Neutrinos from extreme astrophysical sources}}, \href{http://dx.doi.org/10.1016/j.newar.2026.101747}{\emph{New Astron. Rev.} {\bf 102} (2026) 101747}, [\href{http://arxiv.org/abs/2603.10167}{{\tt 2603.10167}}].

\bibitem{Rodrigues:2018tku}
X.~Rodrigues, S.~Gao, A.~Fedynitch, A.~Palladino and W.~Winter, \emph{{Leptohadronic Blazar Models Applied to the 2014{\textendash}2015 Flare of TXS 0506+056}}, \href{http://dx.doi.org/10.3847/2041-8213/ab1267}{\emph{Astrophys. J. Lett.} {\bf 874} (2019) L29}, [\href{http://arxiv.org/abs/1812.05939}{{\tt 1812.05939}}].

\bibitem{Petropoulou:2019zqp}
M.~Petropoulou et~al., \emph{{Multi-Epoch Modeling of TXS 0506+056 and Implications for Long-Term High-Energy Neutrino Emission}}, \href{http://dx.doi.org/10.3847/1538-4357/ab76d0}{\emph{Astrophys. J.} {\bf 891} (2020) 115}, [\href{http://arxiv.org/abs/1911.04010}{{\tt 1911.04010}}].

\bibitem{Murase:2022dog}
K.~Murase, \emph{{Hidden Hearts of Neutrino Active Galaxies}}, \href{http://dx.doi.org/10.3847/2041-8213/aca53c}{\emph{Astrophys. J. Lett.} {\bf 941} (2022) L17}, [\href{http://arxiv.org/abs/2211.04460}{{\tt 2211.04460}}].

\bibitem{Eichmann:2022lxh}
B.~Eichmann, F.~Oikonomou, S.~Salvatore, R.-J. Dettmar and J.~Becker~Tjus, \emph{{Solving the Multimessenger Puzzle of the AGN-starburst Composite Galaxy NGC 1068}}, \href{http://dx.doi.org/10.3847/1538-4357/ac9588}{\emph{Astrophys. J.} {\bf 939} (2022) 43}, [\href{http://arxiv.org/abs/2207.00102}{{\tt 2207.00102}}].

\bibitem{Fiorillo:2023dts}
D.~F.~G. Fiorillo, M.~Petropoulou, L.~Comisso, E.~Peretti and L.~Sironi, \emph{{TeV Neutrinos and Hard X-Rays from Relativistic Reconnection in the Corona of NGC 1068}}, \href{http://dx.doi.org/10.3847/2041-8213/ad192b}{\emph{Astrophys. J.} {\bf 961} (2024) L14}, [\href{http://arxiv.org/abs/2310.18254}{{\tt 2310.18254}}].

\bibitem{Mbarek:2023yeq}
R.~Mbarek, A.~Philippov, A.~Chernoglazov, A.~Levinson and R.~Mushotzky, \emph{{Interplay between accelerated protons, x rays and neutrinos in the corona of NGC 1068: Constraints from kinetic plasma simulations}}, \href{http://dx.doi.org/10.1103/PhysRevD.109.L101306}{\emph{Phys. Rev. D} {\bf 109} (2024) L101306}, [\href{http://arxiv.org/abs/2310.15222}{{\tt 2310.15222}}].

\bibitem{Das:2024vug}
A.~Das, B.~T. Zhang and K.~Murase, \emph{{Revealing the Production Mechanism of High-energy Neutrinos from NGC 1068}}, \href{http://dx.doi.org/10.3847/1538-4357/ad5a04}{\emph{Astrophys. J.} {\bf 972} (2024) 44}, [\href{http://arxiv.org/abs/2405.09332}{{\tt 2405.09332}}].

\bibitem{Das:2026rwt}
A.~Das, K.~Murase and B.~T. Zhang, \emph{{Multimessenger Constraints on Production Sites of High-energy Neutrinos from NGC 1068}}, \href{http://dx.doi.org/10.3847/2041-8213/ae9ba3}{\emph{Astrophys. J. Lett.} {\bf 1009} (2026) L27}, [\href{http://arxiv.org/abs/2604.01222}{{\tt 2604.01222}}].

\bibitem{Kheirandish:2021wkm}
A.~Kheirandish, K.~Murase and S.~S. Kimura, \emph{{High-energy Neutrinos from Magnetized Coronae of Active Galactic Nuclei and Prospects for Identification of Seyfert Galaxies and Quasars in Neutrino Telescopes}}, \href{http://dx.doi.org/10.3847/1538-4357/ac1c77}{\emph{Astrophys. J.} {\bf 922} (2021) 45}, [\href{http://arxiv.org/abs/2102.04475}{{\tt 2102.04475}}].

\bibitem{Murase:2026hrz}
K.~Murase, S.~S. Kimura, M.~Mukhopadhyay and M.~Bhattacharya, \emph{{IceCube's convex all-sky neutrino spectrum consistent with the magnetically powered corona scenario for active galactic nuclei}},  \href{http://arxiv.org/abs/2602.20145}{{\tt 2602.20145}}.

\bibitem{SInoue22}
S.~{Inoue}, M.~{Cerruti}, K.~{Murase} and R.-Y. {Liu}, \emph{{High-energy neutrinos and gamma rays from winds and tori in active galactic nuclei}}, \href{http://dx.doi.org/10.48550/arXiv.2207.02097}{\emph{arXiv e-prints} (July, 2022) arXiv:2207.02097}, [\href{http://arxiv.org/abs/2207.02097}{{\tt 2207.02097}}].

\bibitem{Peretti23}
E.~{Peretti}, A.~{Lamastra}, F.~G. {Saturni}, M.~{Ahlers}, P.~{Blasi}, G.~{Morlino} et~al., \emph{{Diffusive shock acceleration at EeV and associated multimessenger flux from ultra-fast outflows driven by active galactic nuclei}}, \href{http://dx.doi.org/10.1093/mnras/stad2740}{\emph{MNRAS} {\bf 526} (Nov., 2023) 181--192}, [\href{http://arxiv.org/abs/2301.13689}{{\tt 2301.13689}}].

\bibitem{Blanco:2023dfp}
C.~Blanco, D.~Hooper, T.~Linden and E.~Pinetti, \emph{{Neutrino and gamma-ray emissions from NGC 1068}}, \href{http://dx.doi.org/10.1103/wnjh-7nwp}{\emph{Phys. Rev. D} {\bf 112} (2025) 123016}, [\href{http://arxiv.org/abs/2307.03259}{{\tt 2307.03259}}].

\bibitem{Boettcher:2013wxa}
M.~Boettcher, A.~Reimer, K.~Sweeney and A.~Prakash, \emph{{Leptonic and Hadronic Modeling of Fermi-Detected Blazars}}, \href{http://dx.doi.org/10.1088/0004-637X/768/1/54}{\emph{Astrophys. J.} {\bf 768} (2013) 54}, [\href{http://arxiv.org/abs/1304.0605}{{\tt 1304.0605}}].

\bibitem{Murase:2014foa}
K.~Murase, Y.~Inoue and C.~D. Dermer, \emph{{Diffuse Neutrino Intensity from the Inner Jets of Active Galactic Nuclei: Impacts of External Photon Fields and the Blazar Sequence}}, \href{http://dx.doi.org/10.1103/PhysRevD.90.023007}{\emph{Phys. Rev. D} {\bf 90} (2014) 023007}, [\href{http://arxiv.org/abs/1403.4089}{{\tt 1403.4089}}].

\bibitem{Cerruti:2024lmj}
M.~Cerruti et~al., \emph{{A Comprehensive Hadronic Code Comparison for Active Galactic Nuclei}}, \href{http://dx.doi.org/10.3847/1538-4365/ae1d5f}{\emph{Astrophys. J. Suppl.} {\bf 282} (2026) 22}, [\href{http://arxiv.org/abs/2411.14218}{{\tt 2411.14218}}].

\bibitem{Liodakis:2021els}
I.~Liodakis, D.~Blinov, S.~B. Potter and F.~M. Rieger, \emph{{Constraints on magnetic field and particle content in blazar jets through optical circular polarization}}, \href{http://dx.doi.org/10.1093/mnrasl/slab118}{\emph{Mon. Not. Roy. Astron. Soc.} {\bf 509} (2021) L21--L25}, [\href{http://arxiv.org/abs/2110.11434}{{\tt 2110.11434}}].

\bibitem{Reimer:2018vvw}
A.~Reimer, M.~Boettcher and S.~Buson, \emph{{Cascading Constraints from Neutrino-emitting Blazars: The Case of TXS 0506+056}}, \href{http://dx.doi.org/10.3847/1538-4357/ab2bff}{\emph{Astrophys. J.} {\bf 881} (2019) 46}, [\href{http://arxiv.org/abs/1812.05654}{{\tt 1812.05654}}].

\bibitem{Zhang:2019htg}
B.~T. Zhang, M.~Petropoulou, K.~Murase and F.~Oikonomou, \emph{{A Neutral Beam Model for High-Energy Neutrino Emission from the Blazar TXS 0506+056}}, \href{http://dx.doi.org/10.3847/1538-4357/ab659a}{\emph{Astrophys. J.} {\bf 889} (2020) 118}, [\href{http://arxiv.org/abs/1910.11464}{{\tt 1910.11464}}].

\bibitem{Biteau:2020prb}
J.~Biteau, E.~Prandini, L.~Costamante, M.~Lemoine, P.~Padovani, E.~Pueschel et~al., \emph{{Progress in unveiling extreme particle acceleration in persistent astrophysical jets}}, \href{http://dx.doi.org/10.1038/s41550-019-0988-4}{\emph{Nature Astron.} {\bf 4} (2020) 124--131}, [\href{http://arxiv.org/abs/2001.09222}{{\tt 2001.09222}}].

\bibitem{Murase:2023ccp}
K.~Murase, C.~M. Karwin, S.~S. Kimura, M.~Ajello and S.~Buson, \emph{{Sub-GeV Gamma Rays from Nearby Seyfert Galaxies and Implications for Coronal Neutrino Emission}}, \href{http://dx.doi.org/10.3847/2041-8213/ad19c5}{\emph{Astrophys. J. Lett.} {\bf 961} (2024) L34}, [\href{http://arxiv.org/abs/2312.16089}{{\tt 2312.16089}}].

\bibitem{Caputo:2022xpx}
R.~Caputo et~al., \emph{{All-sky Medium Energy Gamma-ray Observatory eXplorer mission concept}}, \href{http://dx.doi.org/10.1117/1.JATIS.8.4.044003}{\emph{J. Astron. Telesc. Instrum. Syst.} {\bf 8} (2022) 044003}, [\href{http://arxiv.org/abs/2208.04990}{{\tt 2208.04990}}].

\bibitem{e-ASTROGAM:2017pxr}
{\scshape e-ASTROGAM} collaboration, M.~Tavani et~al., \emph{{Science with e-ASTROGAM: A space mission for MeV{\textendash}GeV gamma-ray astrophysics}}, \href{http://dx.doi.org/10.1016/j.jheap.2018.07.001}{\emph{JHEAp} {\bf 19} (2018) 1--106}, [\href{http://arxiv.org/abs/1711.01265}{{\tt 1711.01265}}].

\bibitem{Tomsick:2023aue}
J.~A. Tomsick et~al., \emph{{The Compton Spectrometer and Imager}}, \href{http://dx.doi.org/10.22323/1.444.0745}{\emph{PoS} {\bf ICRC2023} (2023) 745}, [\href{http://arxiv.org/abs/2308.12362}{{\tt 2308.12362}}].

\bibitem{Zhang:2019dob}
H.~Zhang, K.~Fang, H.~Li, D.~Giannios, M.~B{\"o}ttcher and S.~Buson, \emph{{Probing the Emission Mechanism and Magnetic Field of Neutrino Blazars with Multiwavelength Polarization Signatures}}, \href{http://dx.doi.org/10.3847/1538-4357/ab158d}{\emph{Astrophys. J.} {\bf 876} (2019) 109}, [\href{http://arxiv.org/abs/1903.01956}{{\tt 1903.01956}}].

\bibitem{Zacharias:2022pea}
M.~Zacharias, A.~Reimer, C.~Boisson and A.~Zech, \emph{{ExHaLe-jet: an extended hadro-leptonic jet model for blazars {\textendash} I. Code description and initial results}}, \href{http://dx.doi.org/10.1093/mnras/stac754}{\emph{Mon. Not. Roy. Astron. Soc.} {\bf 512} (2022) 3948--3971}, [\href{http://arxiv.org/abs/2203.07956}{{\tt 2203.07956}}].

\bibitem{Quinlan:1994ed}
G.~D. Quinlan, L.~Hernquist and S.~Sigurdsson, \emph{{Models of Galaxies with Central Black Holes: Adiabatic Growth in Spherical Galaxies}}, \href{http://dx.doi.org/10.1086/175295}{\emph{Astrophys. J.} {\bf 440} (1995) 554--564}, [\href{http://arxiv.org/abs/astro-ph/9407005}{{\tt astro-ph/9407005}}].

\bibitem{Navarro:1995iw}
J.~F. Navarro, C.~S. Frenk and S.~D.~M. White, \emph{{The Structure of cold dark matter halos}}, \href{http://dx.doi.org/10.1086/177173}{\emph{Astrophys. J.} {\bf 462} (1996) 563--575}, [\href{http://arxiv.org/abs/astro-ph/9508025}{{\tt astro-ph/9508025}}].

\bibitem{Sadeghian:2013laa}
L.~Sadeghian, F.~Ferrer and C.~M. Will, \emph{{Dark matter distributions around massive black holes: A general relativistic analysis}}, \href{http://dx.doi.org/10.1103/PhysRevD.88.063522}{\emph{Phys. Rev. D} {\bf 88} (2013) 063522}, [\href{http://arxiv.org/abs/1305.2619}{{\tt 1305.2619}}].

\bibitem{Merritt:2003}
D.~Merritt, \emph{{Single and Binary Black Holes and their Influence on Nuclear Structure}},  in \emph{{Coevolution of Black Holes and Galaxies}} (L.~C. Ho, ed.), vol.~1 of \emph{Carnegie Observatories Astrophysics Series}.
\newblock Cambridge University Press, Cambridge, 2004.
\newblock \href{http://arxiv.org/abs/astro-ph/0301257}{{\tt astro-ph/0301257}}.

\bibitem{Ullio:2001fb}
P.~Ullio, H.~Zhao and M.~Kamionkowski, \emph{{A Dark matter spike at the galactic center?}}, \href{http://dx.doi.org/10.1103/PhysRevD.64.043504}{\emph{Phys. Rev. D} {\bf 64} (2001) 043504}, [\href{http://arxiv.org/abs/astro-ph/0101481}{{\tt astro-ph/0101481}}].

\bibitem{Merritt:2002vj}
D.~Merritt, M.~Milosavljevic, L.~Verde and R.~Jimenez, \emph{{Dark matter spikes and annihilation radiation from the galactic center}}, \href{http://dx.doi.org/10.1103/PhysRevLett.88.191301}{\emph{Phys. Rev. Lett.} {\bf 88} (2002) 191301}, [\href{http://arxiv.org/abs/astro-ph/0201376}{{\tt astro-ph/0201376}}].

\bibitem{Merritt:2003qk}
D.~Merritt, \emph{{Evolution of the dark matter distribution at the galactic center}}, \href{http://dx.doi.org/10.1103/PhysRevLett.92.201304}{\emph{Phys. Rev. Lett.} {\bf 92} (2004) 201304}, [\href{http://arxiv.org/abs/astro-ph/0311594}{{\tt astro-ph/0311594}}].

\bibitem{Bertone:2005hw}
G.~Bertone and D.~Merritt, \emph{{Time-dependent models for dark matter at the Galactic Center}}, \href{http://dx.doi.org/10.1103/PhysRevD.72.103502}{\emph{Phys. Rev. D} {\bf 72} (2005) 103502}, [\href{http://arxiv.org/abs/astro-ph/0501555}{{\tt astro-ph/0501555}}].

\bibitem{Vasiliev:2008uz}
E.~Vasiliev and M.~Zelnikov, \emph{{Dark matter dynamics in Galactic center}}, \href{http://dx.doi.org/10.1103/PhysRevD.78.083506}{\emph{Phys. Rev. D} {\bf 78} (2008) 083506}, [\href{http://arxiv.org/abs/0803.0002}{{\tt 0803.0002}}].

\bibitem{Ferrer:2017xwm}
F.~Ferrer, A.~M. da~Rosa and C.~M. Will, \emph{{Dark matter spikes in the vicinity of Kerr black holes}}, \href{http://dx.doi.org/10.1103/PhysRevD.96.083014}{\emph{Phys. Rev. D} {\bf 96} (2017) 083014}, [\href{http://arxiv.org/abs/1707.06302}{{\tt 1707.06302}}].

\bibitem{Shapiro:2022prq}
S.~L. Shapiro and D.~C. Heggie, \emph{{Effect of stars on the dark matter spike around a black hole: A tale of two treatments}}, \href{http://dx.doi.org/10.1103/PhysRevD.106.043018}{\emph{Phys. Rev. D} {\bf 106} (2022) 043018}, [\href{http://arxiv.org/abs/2209.08105}{{\tt 2209.08105}}].

\bibitem{Bertone:2024wbn}
G.~Bertone, A.~R. A.~C. Wierda, D.~Gaggero, B.~J. Kavanagh, M.~Volonteri and N.~Yoshida, \emph{{Toward a realistic description of dark matter overdensities around black holes}}, \href{http://dx.doi.org/10.1103/5nnf-8fz9}{\emph{Phys. Rev. D} {\bf 112} (2025) 043537}, [\href{http://arxiv.org/abs/2404.08731}{{\tt 2404.08731}}].

\bibitem{Caiozzo:2025mye}
R.~Caiozzo, G.~Bertone, P.~Ullio, R.~Vicente, B.~J. Kavanagh and D.~Gaggero, \emph{{Dark matter mounds from the collapse of supermassive stars: A general-relativistic analysis}}, \href{http://dx.doi.org/10.1103/lnhp-tffg}{\emph{Phys. Rev. D} {\bf 113} (2026) 063052}, [\href{http://arxiv.org/abs/2512.09985}{{\tt 2512.09985}}].

\bibitem{Sharpe:2026nqq}
C.~Sharpe, Y.~B. Ginat, T.~F.~M. Spieksma and B.~Kocsis, \emph{{The Depletion of Collisionless Dark Matter Spikes}},  \href{http://arxiv.org/abs/2603.28866}{{\tt 2603.28866}}.

\bibitem{Herrera:2026hal}
G.~Herrera, A.~Hussein, L.~Necib, E.~Y. Davies and X.~Shen, \emph{{Formation and Redshift Evolution of Dark Matter Spikes}},  \href{http://arxiv.org/abs/2605.01023}{{\tt 2605.01023}}.

\bibitem{Karydas:2026ctq}
T.~K. Karydas, F.~Scarcella, B.~J. Kavanagh and G.~Bertone, \emph{{On the survival of dark matter spikes: Stellar and compact-object perturbations}},  \href{http://arxiv.org/abs/2606.13761}{{\tt 2606.13761}}.

\bibitem{Sharpe:2026kfk}
C.~Sharpe, Y.~B. Ginat, T.~F.~M. Spieksma, A.~Heger and B.~Kocsis, \emph{{The Irreversible Depletion of Dark Matter Spikes by Extreme-Mass-Ratio Inspirals}},  \href{http://arxiv.org/abs/2609.37234}{{\tt 2609.37234}}.

\bibitem{Belikov:2013nca}
A.~Belikov and J.~Silk, \emph{{Diffuse Gamma Ray Background from Annihilating Dark Matter in Density Spikes around Supermassive Black Holes}}, \href{http://dx.doi.org/10.1103/PhysRevD.89.043520}{\emph{Phys. Rev. D} {\bf 89} (2014) 043520}, [\href{http://arxiv.org/abs/1312.0007}{{\tt 1312.0007}}].

\bibitem{Lacroix:2015lxa}
T.~Lacroix, C.~B{\oe}hm and J.~Silk, \emph{{Ruling out thermal dark matter with a black hole induced spiky profile in the M87 galaxy}}, \href{http://dx.doi.org/10.1103/PhysRevD.92.043510}{\emph{Phys. Rev. D} {\bf 92} (2015) 043510}, [\href{http://arxiv.org/abs/1505.00785}{{\tt 1505.00785}}].

\bibitem{Shelton:2015aqa}
J.~Shelton, S.~L. Shapiro and B.~D. Fields, \emph{{Black hole window into $p$-wave dark matter annihilation}}, \href{http://dx.doi.org/10.1103/PhysRevLett.115.231302}{\emph{Phys. Rev. Lett.} {\bf 115} (2015) 231302}, [\href{http://arxiv.org/abs/1506.04143}{{\tt 1506.04143}}].

\bibitem{Eda:2013gg}
K.~Eda, Y.~Itoh, S.~Kuroyanagi and J.~Silk, \emph{{New Probe of Dark-Matter Properties: Gravitational Waves from an Intermediate-Mass Black Hole Embedded in a Dark-Matter Minispike}}, \href{http://dx.doi.org/10.1103/PhysRevLett.110.221101}{\emph{Phys. Rev. Lett.} {\bf 110} (2013) 221101}, [\href{http://arxiv.org/abs/1301.5971}{{\tt 1301.5971}}].

\bibitem{Eda:2014kra}
K.~Eda, Y.~Itoh, S.~Kuroyanagi and J.~Silk, \emph{{Gravitational waves as a probe of dark matter minispikes}}, \href{http://dx.doi.org/10.1103/PhysRevD.91.044045}{\emph{Phys. Rev. D} {\bf 91} (2015) 044045}, [\href{http://arxiv.org/abs/1408.3534}{{\tt 1408.3534}}].

\bibitem{Macedo:2013qea}
C.~F.~B. Macedo, P.~Pani, V.~Cardoso and L.~C.~B. Crispino, \emph{{Into the lair: gravitational-wave signatures of dark matter}}, \href{http://dx.doi.org/10.1088/0004-637X/774/1/48}{\emph{Astrophys. J.} {\bf 774} (2013) 48}, [\href{http://arxiv.org/abs/1302.2646}{{\tt 1302.2646}}].

\bibitem{Barausse:2014tra}
E.~Barausse, V.~Cardoso and P.~Pani, \emph{{Can environmental effects spoil precision gravitational-wave astrophysics?}}, \href{http://dx.doi.org/10.1103/PhysRevD.89.104059}{\emph{Phys. Rev. D} {\bf 89} (2014) 104059}, [\href{http://arxiv.org/abs/1404.7149}{{\tt 1404.7149}}].

\bibitem{Karydas:2024fcn}
T.~K. Karydas, B.~J. Kavanagh and G.~Bertone, \emph{{Sharpening the dark matter signature in gravitational waveforms.~I.~Accretion and eccentricity evolution}}, \href{http://dx.doi.org/10.1103/PhysRevD.111.063070}{\emph{Phys. Rev. D} {\bf 111} (2025) 063070}, [\href{http://arxiv.org/abs/2402.13053}{{\tt 2402.13053}}].

\bibitem{Kavanagh:2024lgq}
B.~J. Kavanagh, T.~K. Karydas, G.~Bertone, P.~Di~Cintio and M.~Pasquato, \emph{{Sharpening the dark matter signature in gravitational waveforms. II. Numerical simulations}}, \href{http://dx.doi.org/10.1103/PhysRevD.111.063071}{\emph{Phys. Rev. D} {\bf 111} (2025) 063071}, [\href{http://arxiv.org/abs/2402.13762}{{\tt 2402.13762}}].

\bibitem{Sharma:2025ynw}
M.~Sharma, G.~Herrera, N.~Arav and S.~Horiuchi, \emph{{Novel method to trace the dark matter density profile around supermassive black holes with AGN reverberation mapping}}, \href{http://dx.doi.org/10.1103/llpr-gnmh}{\emph{Phys. Rev. D} {\bf 113} (2026) 043052}, [\href{http://arxiv.org/abs/2506.10122}{{\tt 2506.10122}}].

\bibitem{Goerdt:2008}
T.~Goerdt, B.~Moore, J.~I. Read and J.~Stadel, \emph{{Core Creation in Galaxies and Haloes Via Sinking Massive Objects}}, \href{http://dx.doi.org/10.1088/0004-637X/725/2/1707}{\emph{Astrophys. J.} {\bf 725} (2010) 1707--1716}, [\href{http://arxiv.org/abs/0806.1951}{{\tt 0806.1951}}].

\bibitem{Merritt:2006mt}
D.~Merritt, S.~Harfst and G.~Bertone, \emph{{Collisionally Regenerated Dark Matter Structures in Galactic Nuclei}}, \href{http://dx.doi.org/10.1103/PhysRevD.75.043517}{\emph{Phys. Rev. D} {\bf 75} (2007) 043517}, [\href{http://arxiv.org/abs/astro-ph/0610425}{{\tt astro-ph/0610425}}].

\bibitem{Kavanagh:2020cfn}
B.~J. Kavanagh, D.~A. Nichols, G.~Bertone and D.~Gaggero, \emph{{Detecting dark matter around black holes with gravitational waves: Effects of dark-matter dynamics on the gravitational waveform}}, \href{http://dx.doi.org/10.1103/PhysRevD.102.083006}{\emph{Phys. Rev. D} {\bf 102} (2020) 083006}, [\href{http://arxiv.org/abs/2002.12811}{{\tt 2002.12811}}].

\bibitem{Kavanagh:2018}
B.~J. Kavanagh, D.~Gaggero and G.~Bertone, \emph{{Merger rate of a subdominant population of primordial black holes}}, \href{http://dx.doi.org/10.1103/PhysRevD.98.023536}{\emph{Phys. Rev. D} {\bf 98} (2018) 023536}, [\href{http://arxiv.org/abs/1805.09034}{{\tt 1805.09034}}].

\bibitem{Vasiliev:2007}
E.~Vasiliev, \emph{{Dark matter annihilation near a black hole: Plateau versus weak cusp}}, \href{http://dx.doi.org/10.1103/PhysRevD.76.103532}{\emph{Phys. Rev. D} {\bf 76} (2007) 103532}, [\href{http://arxiv.org/abs/0707.3334}{{\tt 0707.3334}}].

\bibitem{ShapiroShelton:2016}
S.~L. Shapiro and J.~Shelton, \emph{{Weak annihilation cusp inside the dark matter spike about a black hole}}, \href{http://dx.doi.org/10.1103/PhysRevD.93.123510}{\emph{Phys. Rev. D} {\bf 93} (2016) 123510}, [\href{http://arxiv.org/abs/1606.01248}{{\tt 1606.01248}}].

\bibitem{Shapiro:2018}
S.~L. Shapiro, \emph{{Star clusters, self-interacting dark matter halos, and black hole cusps: The fluid conduction model and its extension to general relativity}}, \href{http://dx.doi.org/10.1103/PhysRevD.98.023021}{\emph{Phys. Rev. D} {\bf 98} (2018) 023021}, [\href{http://arxiv.org/abs/1809.02618}{{\tt 1809.02618}}].

\bibitem{ShapiroPaschalidis:2014}
S.~L. Shapiro and V.~Paschalidis, \emph{{Self-interacting dark matter cusps around massive black holes}}, \href{http://dx.doi.org/10.1103/PhysRevD.89.023506}{\emph{Phys. Rev. D} {\bf 89} (2014) 023506}, [\href{http://arxiv.org/abs/1402.0005}{{\tt 1402.0005}}].

\bibitem{2004ApJ...613L.109P}
M.~{Preto}, D.~{Merritt} and R.~{Spurzem}, \emph{{N-Body Growth of a Bahcall-Wolf Cusp around a Black Hole}}, \href{http://dx.doi.org/10.1086/425139}{\emph{ApJL} {\bf 613} (Oct., 2004) L109--L112}, [\href{http://arxiv.org/abs/astro-ph/0406324}{{\tt astro-ph/0406324}}].

\bibitem{2023A&A...673A...8D}
P.~{Di Cintio}, M.~{Pasquato}, L.~{Barbieri}, A.~A. {Trani} and U.~N. {di Carlo}, \emph{{Dynamics of intermediate mass black holes in globular clusters. Wander radius and anisotropy profiles}}, \href{http://dx.doi.org/10.1051/0004-6361/202346124}{\emph{A\&A} {\bf 673} (May, 2023) A8}, [\href{http://arxiv.org/abs/2302.05542}{{\tt 2302.05542}}].

\bibitem{2025A&A...700A.230D}
P.~{Di Cintio} and B.~{Marcos}, \emph{{A disturbance in the force: How force fluctuations hinder dynamical friction and induce core stalling}}, \href{http://dx.doi.org/10.1051/0004-6361/202555427}{\emph{A\&A} {\bf 700} (Aug., 2025) A230}, [\href{http://arxiv.org/abs/2505.04505}{{\tt 2505.04505}}].

\bibitem{2021ApJ...912...43B}
U.~{Banik} and F.~C. {van den Bosch}, \emph{{A Self-consistent, Time-dependent Treatment of Dynamical Friction: New Insights Regarding Core Stalling and Dynamical Buoyancy}}, \href{http://dx.doi.org/10.3847/1538-4357/abeb6d}{\emph{ApJ} {\bf 912} (May, 2021) 43}, [\href{http://arxiv.org/abs/2103.05004}{{\tt 2103.05004}}].

\bibitem{2022ApJ...926..215B}
U.~{Banik} and F.~C. {van den Bosch}, \emph{{Dynamical Friction, Buoyancy, and Core-stalling. I. A Nonperturbative Orbit-based Analysis}}, \href{http://dx.doi.org/10.3847/1538-4357/ac4242}{\emph{ApJ} {\bf 926} (Feb., 2022) 215}, [\href{http://arxiv.org/abs/2112.06944}{{\tt 2112.06944}}].

\bibitem{2025arXiv251111804D}
S.~{Dattathri}, F.~C. {van den Bosch}, U.~{Banik}, M.~{Weinberg}, P.~{Natarajan}, Z.~{Li} et~al., \emph{{Not all cores are equal: Phase-space origins of dynamical friction, stalling and buoyancy}}, \href{http://dx.doi.org/10.48550/arXiv.2511.11804}{\emph{arXiv e-prints} (Nov., 2025) arXiv:2511.11804}, [\href{http://arxiv.org/abs/2511.11804}{{\tt 2511.11804}}].

\bibitem{Kormendy2013}
J.~{Kormendy} and L.~C. {Ho}, \emph{{Coevolution (Or Not) of Supermassive Black Holes and Host Galaxies}}, \href{http://dx.doi.org/10.1146/annurev-astro-082708-101811}{\emph{Annual Review of Astronomy and Astrophysics} {\bf 51} (Aug., 2013) 511--653}, [\href{http://arxiv.org/abs/1304.7762}{{\tt 1304.7762}}].

\bibitem{Rosenthal2020}
M.~J. {Rosenthal} and I.~{Zaw}, \emph{{A search for H$_{2}$O masers in 100 active dwarf galaxies}}, \href{http://dx.doi.org/10.1093/mnras/staa2730}{\emph{MNRAS} {\bf 499} (Nov., 2020) 1233--1249}, [\href{http://arxiv.org/abs/2009.01104}{{\tt 2009.01104}}].

\bibitem{Lo2005}
K.~Y. {Lo}, \emph{{Mega-Masers and Galaxies}}, \href{http://dx.doi.org/10.1146/annurev.astro.41.011802.094927}{\emph{Annual Review of Astronomy and Astrophysics} {\bf 43} (Sept., 2005) 625--676}.

\bibitem{Blandford1982}
R.~D. {Blandford} and C.~F. {McKee}, \emph{{Reverberation mapping of the emission line regions of Seyfert galaxies and quasars.}}, \href{http://dx.doi.org/10.1086/159843}{\emph{ApJ} {\bf 255} (Apr., 1982) 419--439}.

\bibitem{Peterson1993}
B.~M. {Peterson}, \emph{{Reverberation Mapping of Active Galactic Nuclei}}, \href{http://dx.doi.org/10.1086/133140}{\emph{PASP} {\bf 105} (Mar., 1993) 247}.

\bibitem{Shen2024}
Y.~{Shen}, C.~J. {Grier}, K.~{Horne}, Z.~{Stone}, J.~I. {Li}, Q.~{Yang} et~al., \emph{{The Sloan Digital Sky Survey Reverberation Mapping Project: Key Results}}, \href{http://dx.doi.org/10.3847/1538-4365/ad3936}{\emph{ApJs} {\bf 272} (June, 2024) 26}, [\href{http://arxiv.org/abs/2305.01014}{{\tt 2305.01014}}].

\bibitem{Cackett2021}
E.~M. {Cackett}, M.~C. {Bentz} and E.~{Kara}, \emph{{Reverberation mapping of active galactic nuclei: from X-ray corona to dusty torus}}, \href{http://dx.doi.org/10.1016/j.isci.2021.102557}{\emph{iScience} {\bf 24} (June, 2021) 102557}, [\href{http://arxiv.org/abs/2105.06926}{{\tt 2105.06926}}].

\bibitem{Horne2004}
K.~{Horne}, B.~M. {Peterson}, S.~J. {Collier} and H.~{Netzer}, \emph{{Observational Requirements for High-Fidelity Reverberation Mapping}}, \href{http://dx.doi.org/10.1086/420755}{\emph{PASP} {\bf 116} (May, 2004) 465--476}.

\bibitem{Onken2004}
C.~A. {Onken}, L.~{Ferrarese}, D.~{Merritt}, B.~M. {Peterson}, R.~W. {Pogge}, M.~{Vestergaard} et~al., \emph{{Supermassive Black Holes in Active Galactic Nuclei. II. Calibration of the Black Hole Mass-Velocity Dispersion Relationship for Active Galactic Nuclei}}, \href{http://dx.doi.org/10.1086/424655}{\emph{ApJ} {\bf 615} (Nov., 2004) 645--651}, [\href{http://arxiv.org/abs/astro-ph/0407297}{{\tt astro-ph/0407297}}].

\bibitem{Batiste2017}
M.~{Batiste}, M.~C. {Bentz}, S.~I. {Raimundo}, M.~{Vestergaard} and C.~A. {Onken}, \emph{{Recalibration of the M $_{BH}$-{\ensuremath{\sigma}} $_{{\ensuremath{\star}}}$ Relation for AGN}}, \href{http://dx.doi.org/10.3847/2041-8213/aa6571}{\emph{ApJl} {\bf 838} (Mar., 2017) L10}, [\href{http://arxiv.org/abs/1612.02815}{{\tt 1612.02815}}].

\bibitem{DeRosa2015}
G.~{De Rosa}, B.~M. {Peterson}, J.~{Ely}, G.~A. {Kriss}, D.~M. {Crenshaw}, K.~{Horne} et~al., \emph{{Space Telescope and Optical Reverberation Mapping Project.I. Ultraviolet Observations of the Seyfert 1 Galaxy NGC 5548 with the Cosmic Origins Spectrograph on Hubble Space Telescope}}, \href{http://dx.doi.org/10.1088/0004-637X/806/1/128}{\emph{ApJ} {\bf 806} (June, 2015) 128}, [\href{http://arxiv.org/abs/1501.05954}{{\tt 1501.05954}}].

\bibitem{Williams2020}
P.~R. {Williams}, A.~{Pancoast}, T.~{Treu}, B.~J. {Brewer}, B.~M. {Peterson}, A.~J. {Barth} et~al., \emph{{Space Telescope and Optical Reverberation Mapping Project. XII. Broad-line Region Modeling of NGC 5548}}, \href{http://dx.doi.org/10.3847/1538-4357/abbad7}{\emph{ApJ} {\bf 902} (Oct., 2020) 74}, [\href{http://arxiv.org/abs/2010.00594}{{\tt 2010.00594}}].

\bibitem{Homayouni2023}
Y.~{Homayouni}, G.~{De Rosa}, R.~{Plesha}, G.~A. {Kriss}, A.~J. {Barth}, E.~M. {Cackett} et~al., \emph{{AGN STORM 2. II. Ultraviolet Observations of Mrk 817 with the Cosmic Origins Spectrograph on the Hubble Space Telescope}}, \href{http://dx.doi.org/10.3847/1538-4357/acc45a}{\emph{ApJ} {\bf 948} (May, 2023) 85}, [\href{http://arxiv.org/abs/2302.11587}{{\tt 2302.11587}}].

\bibitem{Bentz2021}
M.~C. {Bentz}, R.~{Street}, C.~A. {Onken} and M.~{Valluri}, \emph{{Robotic Reverberation Mapping of the Southern Seyfert NGC 3783}}, \href{http://dx.doi.org/10.3847/1538-4357/abccd4}{\emph{ApJ} {\bf 906} (Jan., 2021) 50}, [\href{http://arxiv.org/abs/2009.11358}{{\tt 2009.11358}}].

\bibitem{Bentz2021a}
M.~C. {Bentz}, P.~R. {Williams}, R.~{Street}, C.~A. {Onken}, M.~{Valluri} and T.~{Treu}, \emph{{A Detailed View of the Broad-line Region in NGC 3783 from Velocity-resolved Reverberation Mapping}}, \href{http://dx.doi.org/10.3847/1538-4357/ac19af}{\emph{ApJ} {\bf 920} (Oct., 2021) 112}, [\href{http://arxiv.org/abs/2108.00482}{{\tt 2108.00482}}].

\bibitem{Bentz2010}
M.~C. {Bentz}, K.~{Horne}, A.~J. {Barth}, V.~N. {Bennert}, G.~{Canalizo}, A.~V. {Filippenko} et~al., \emph{{The Lick AGN Monitoring Project: Velocity-delay Maps from the Maximum-entropy Method for Arp 151}}, \href{http://dx.doi.org/10.1088/2041-8205/720/1/L46}{\emph{ApJl} {\bf 720} (Sept., 2010) L46--L51}, [\href{http://arxiv.org/abs/1007.0781}{{\tt 1007.0781}}].

\bibitem{Rosborough2025}
S.~A. {Rosborough}, A.~{Robinson}, T.~{Almeyda}, D.~{Humphrey} and M.~{Noll}, \emph{{Modeling the Reverberation Response of the Broad-line Region in Active Galactic Nuclei. II. Incorporating Photoionization Models}}, \href{http://dx.doi.org/10.3847/1538-4357/adf109}{\emph{ApJ} {\bf 991} (Oct., 2025) 143}, [\href{http://arxiv.org/abs/2507.14337}{{\tt 2507.14337}}].

\bibitem{Herrera:2026mao}
G.~Herrera, \emph{{Population search for dark matter spikes in megamaser rotation curves}},  \href{http://arxiv.org/abs/2609.36729}{{\tt 2609.36729}}.

\bibitem{Bertone:2019irm}
G.~Bertone et~al., \emph{{Gravitational wave probes of dark matter: challenges and opportunities}}, \href{http://dx.doi.org/10.21468/SciPostPhysCore.3.2.007}{\emph{SciPost Phys. Core} {\bf 3} (2020) 007}, [\href{http://arxiv.org/abs/1907.10610}{{\tt 1907.10610}}].

\bibitem{Cole:2022yzw}
P.~S. Cole, G.~Bertone, A.~Coogan, D.~Gaggero, T.~Karydas, B.~J. Kavanagh et~al., \emph{{Distinguishing environmental effects on binary black hole gravitational waveforms}}, \href{http://dx.doi.org/10.1038/s41550-023-01990-2}{\emph{Nature Astron.} {\bf 7} (2023) 943--950}, [\href{http://arxiv.org/abs/2211.01362}{{\tt 2211.01362}}].

\bibitem{DiCintio:2025ilt}
P.~{Di Cintio}, G.~{Bertone}, C.~{Chiari}, T.~K. {Karydas}, B.~J. {Kavanagh}, M.~{Pasquato} et~al., \emph{{GW emission and relativistic dynamical friction in intermediate mass ratio inspirals}},  in \emph{IAU Symposium} (H.~M. {Lee}, J.~{Hong} and R.~{Spruzem}, eds.), vol.~21 of \emph{IAU Symposium}, pp.~294--298, CUP, Sept., 2025.
\newblock \href{http://dx.doi.org/10.1017/S1743921325104924}{DOI}.

\bibitem{Vicente:2025gsg}
R.~Vicente, T.~K. Karydas and G.~Bertone, \emph{{Fully Relativistic Treatment of Extreme Mass-Ratio Inspirals in Collisionless Environments}}, \href{http://dx.doi.org/10.1103/s4wh-x6c4}{\emph{Phys. Rev. Lett.} {\bf 135} (2025) 211401}, [\href{http://arxiv.org/abs/2505.09715}{{\tt 2505.09715}}].

\bibitem{Alachkar:2022oj287}
A.~Alachkar, J.~Ellis and M.~Fairbairn, \emph{{Dark Matter Constraints from the Eccentric Supermassive Black Hole Binary OJ 287}}, \href{http://dx.doi.org/10.1103/PhysRevD.107.103033}{\emph{Phys. Rev. D} {\bf 107} (2023) 103033}, [\href{http://arxiv.org/abs/2207.10021}{{\tt 2207.10021}}].

\bibitem{Chan:2024oj287}
M.~H. Chan and C.~M. Lee, \emph{{The First Robust Evidence Showing a Dark Matter Density Spike around the Supermassive Black Hole in OJ 287}}, {\emph{Astrophys. J. Lett.} {\bf 962} (2024) L40}, [\href{http://arxiv.org/abs/2402.03751}{{\tt 2402.03751}}].

\bibitem{Deb:2025raq}
D.~Deb, A.~Gopakumar and M.~J. Valtonen, \emph{{On the Evidence of a Dark Matter Density Spike around the Primary Black Hole in OJ 287}}, \href{http://dx.doi.org/10.3847/2041-8213/adcee3}{\emph{Astrophys. J. Lett.} {\bf 984} (2025) L66}, [\href{http://arxiv.org/abs/2504.05715}{{\tt 2504.05715}}].

\bibitem{GRAVITY:2024tth}
{\scshape GRAVITY} collaboration, K.~Abd El~Dayem et~al., \emph{{Improving constraints on the extended mass distribution in the Galactic center with stellar orbits}}, \href{http://dx.doi.org/10.1051/0004-6361/202452274}{\emph{Astron. Astrophys.} {\bf 692} (2024) A242}, [\href{http://arxiv.org/abs/2409.12261}{{\tt 2409.12261}}].

\bibitem{Antonini:2011tu}
F.~Antonini and D.~Merritt, \emph{{Dynamical Friction around Supermassive Black Holes}}, \href{http://dx.doi.org/10.1088/0004-637X/745/1/83}{\emph{Astrophys. J.} {\bf 745} (2012) 83}, [\href{http://arxiv.org/abs/1108.1163}{{\tt 1108.1163}}].

\bibitem{Dosopoulou:2023umg}
F.~Dosopoulou, \emph{{Dynamical friction in dark matter spikes: Corrections to Chandrasekhar{\textquoteright}s formula}}, \href{http://dx.doi.org/10.1103/PhysRevD.110.083027}{\emph{Phys. Rev. D} {\bf 110} (2024) 083027}, [\href{http://arxiv.org/abs/2305.17281}{{\tt 2305.17281}}].

\bibitem{Gustafson:2025ypo}
R.~A. Gustafson, I.~M. Shoemaker and V.~Takhistov, \emph{{Probing Dark Matter Interactions with Stellar Motion near Sagittarius A*}},  \href{http://arxiv.org/abs/2510.07387}{{\tt 2510.07387}}.

\bibitem{Acevedo:2025rqu}
J.~F. Acevedo, A.~J. Reilly and L.~Santos-Olmsted, \emph{{Dark Drag Around Sagittarius A*}},  \href{http://arxiv.org/abs/2510.01320}{{\tt 2510.01320}}.

\end{thebibliography}\endgroup

% end of \small{}
}
\end{document}